\pdfoutput=1
\documentclass[12pt,reqno]{article}
\usepackage{feynmp-auto}
\usepackage{jheppub}
\usepackage{amsmath,amssymb,amsfonts}
\graphicspath{ {./Graphs/} }
\usepackage{float}             
 \usepackage[T1]{fontenc}
\usepackage[utf8]{inputenc}
\usepackage{soul}
\usepackage{mathtools}  
\usepackage{braket}
\usepackage{blkarray}
\usepackage[usenames,dvipsnames]{xcolor}
\usepackage{epsfig}
\usepackage{epstopdf}
\usepackage{dcolumn}
\usepackage{tikz}
\usetikzlibrary{shapes.geometric, arrows}
\usepackage{upgreek}
\usepackage{setspace}
\usepackage{enumitem}
\usepackage{array,multirow,bigdelim}
\usepackage{cleveref}
\usepackage{bm}

\def\be{\begin{equation}}
\def\ee{\end{equation}}
\def\ba{\begin{eqnarray}}
\def\ea{\end{eqnarray}}

\def\CP1{\mathbb{CP}^1}
\def\SL2C{\mathrm{SL}(2,\mathbb{C})}

\def\Z2{\mathbb{Z}_2}

\def\su2{{SU(2)}}

\def\[{\left[}
\def\]{\right]}

\def\({\left(}
\def\){\right)}
\def\[{\left[}
\def\]{\right]}

\def\<{\langle}
\def\>{\rangle}

\def\i2{\frac{i}{2}}

\newcommand{\bea}{\begin{eqnarray}}
\newcommand{\eea}{\end{eqnarray}}
\newcommand{\beq}{\begin{equation}}
\newcommand{\eeq}{\end{equation}}
\newcommand{\bean}{\begin{eqnarray*}}
\newcommand{\eean}{\end{eqnarray*}}

\renewcommand{\tilde}{\widetilde}

\renewcommand{\phi}{\varphi}

\renewcommand{\phi}{\varphi}

\preprint{
	\vspace{-26pt}
\begin{flushright}
	SAGEX-20-05-E\\
\hfill \hspace{10pt} NORDITA 2020-026
\end{flushright}
}
\title{Second-order Post-Minkowskian scattering in arbitrary dimensions} 
\author[1]{Andrea~Cristofoli}
\author[1]{Poul~H.~Damgaard}
\author[1,2]{Paolo~Di~Vecchia}
\author[2]{Carlo~Heissenberg}
\affiliation[1]{Niels Bohr International Academy and Discovery Center\\ 
The Niels Bohr Institute, University of Copenhagen\\
Blegdamsvej 17, DK-2100 Copenhagen, Denmark}
\affiliation[2]{NORDITA, KTH Royal Institute of Technology and Stockholm University\\
Roslagstullsbacken 23,SE-10691 Stockholm, Sweden}
\emailAdd{a.cristofoli@nbi.ku.dk}
\emailAdd{phdamg@nbi.dk}
\emailAdd{divecchi@nbi.dk}
\emailAdd{carlo.heissenberg@su.se}
\keywords{Scattering Amplitudes, General Relativity}
\date{\today}

\abstract{
	We extract the long-range gravitational potential between two scalar particles with arbitrary masses from the two-to-two elastic scattering amplitude at 2nd Post-Minkowskian order in arbitrary dimensions.  
	In contrast to the four-dimensional case, in higher dimensions the classical potential receives contributions from box topologies.
	Moreover, the kinematical relation between momentum and position on the classical trajectory contains a new term which is quadratic in the tree-level amplitude. 
	A precise interplay between this new relation and the formula for the scattering angle  ensures that the latter is still linear in the classical part of the scattering amplitude, to this order, matching an earlier calculation in the eikonal approach. 
	We point out that both the eikonal exponentiation and the reality of the potential to 2nd post-Minkowskian order can be seen as a consequence of unitarity. 
	We finally present closed-form expressions for the scattering angle given by leading-order gravitational potentials for dimensions ranging from four to ten.}

\begin{document}
\maketitle

\section{Introduction}\label{sec:introduction}

The study of gravitational collisions has recently received a lot of attention thanks to the amazing experimental breakthroughs in the detection~\cite{Abbott:2016blz,Abbott:2016nmj,Abbott:2017vtc,Abbott:2017oio,TheLIGOScientific:2017qsa} of gravitational-waves coming from  black-hole or neutron star mergers.  Given  the expected improvements in detector sensitivity, it will be extremely important in the future to have high-precision theoretical predictions from General Relativity.  
To this aim the use of quantum field theory amplitudes to extract the post-Minkowskian (PM)
expansion of General Relativity has recently gained considerable momentum 
\cite
{Damouruno,Damourdue,Bjerrum-Bohruno,Cheung,Bernuno,Antonelli,Cristofoliuno,Berndue,Kalin,Cristofolidue,Blumleinuno,Damourtre,Damourquatro,Blumlein:2020znm,Cheung:2020gyp,Bini:2020wpo},
and progress is now also being made on extensions to spinning objects
\cite{Vinesuno,Vinesdue,Guevarauno,Chung,Maybeeuno,Guevaradue,Nima,Damgaard,Vinestre,Aoude}. The underlying physical motivation for this approach lies in the observation that, during the early stages of a merger event, when the two compact objects are still far apart, gravitational interactions are weak and can be conveniently treated in a weak-coupling approximation. The perturbative series that naturally organizes the calculation of scattering amplitudes in quantum field theory then offers a convenient tool to study the dynamics of such systems for weak gravitational fields without the need to consider the limit of small velocities, thanks to the Lorentz invariance of the amplitude. The price one has to pay in order to eventually retrieve predictions for General Relativity is the proper handling of the classical limit. Indeed, going to higher orders in the gravitational coupling in the classical theory entails evaluating Feynman diagrams with more and more loops in the quantum theory and one may wonder as to how the loop expansion may yield precision corrections to classical quantities, an issue that was first clarified in the seminal papers \cite{Iwasaki:1971vb, Holstein:2004dn} and more recently investigated systematically in \cite{Kosower}.

A fundamental and gauge-independent quantity that is most readily computed from quantum field theoretic amplitudes is the scattering angle of two colliding massive objects. 
Computations of classical gravitational observables using relativistic amplitude techniques have so far been performed with two 
\emph{a~priori} different approaches. One is based on the evaluation of the eikonal phase, while the other proceeds by constructing the Hamiltonian, \emph{i.e.} the effective interaction potential. The deflection angle can then be easily obtained from either of these two quantities.   

The eikonal approach began in the late eighties with the work by 't Hooft~\cite{'tHooft} and independent parallel work of two other groups~\cite{Amati:1987wq,Muzinich:1987in,Amati:1987uf}, dealing with transplanckian energy collisions of strings in a generic number $D$ of macroscopic  dimensions.  
It was further developed in Refs.~\cite{Amati:1990xe,Verlinde:1991iu,Amati:1992zb,Bellini:1992eb,Amati:1993tb,Giddings:2010pp,Kabat:1992tb,Bjerrum-Bohruno,Sterman,Luna,Collado:2018isu,KoemansCollado:2019lnh,Paolodue} and  extended to the scattering of strings off a stack of
$D$-branes~\cite{DAppollonio:2010krb,DAppollonio:2013mgj} (see Ref.~\cite{Veneziano:2015kfb} for a review) and recently to supersymmetric theories~\cite{Paolotre,Paoloquatro,Berntre}. 

That approach has its origin in the observation that, in general, a tree diagram in gravity diverges at high energy, implying that unitarity is violated in this regime. A viable way to restore unitarity is to first observe that also the loop diagrams are  divergent at high energy and actually their degree of divergence increases with the number of loops. Then, Fourier transforming a  suitably normalized amplitude from momentum space to the $(D-2)$-dimensional impact parameter space, 
one sees that the leading terms for large impact parameter of the various diagrams re-sum into an exponential given by the tree contribution,  whose phase is called the leading eikonal. In this way one obtains a quantity that is consistent with unitarity. 
Sub-leading eikonals can be obtained in a similar way by re-summing 
diagrams that are subdominant for large impact parameter. Unlike the leading one, they also contain an imaginary part related to inelastic processes, although we do not discuss these new effects in this paper.  

Having determined the eikonal, one can then use it to compute the classical deflection angle taking its derivative with respect to the impact parameter.  Other physical quantities, as for instance the Shapiro time delay, can also be computed  from the eikonal. An interesting aspect of this approach  is that, in order to compute the deflection angle to a given order in the coupling, one must still compute, in principle, an infinite number of loops to check the exponentiation. 

In contrast, the Hamiltonian approach relies on the calculation of the effective interaction potential between two massive particles from the scattering amplitude, which is achieved as follows. 
One first imposes that the two-to-two scattering amplitude in General Relativity be equal to that of an effective theory of massive particles interacting via a long-range potential and then reconstructs the potential that ensures this matching condition order by order in Newton's coupling constant $G_N$. 
To this purpose one can either employ the relativistic Lippmann--Schwinger equation and the technique of Born subtractions for a first-quantized effective theory \cite{Cristofoliuno, Cristofolidue}, or alternatively the Effective Field Theory (EFT) matching procedure \cite{Cheung, Berndue}. 

These two methods have proven to be completely equivalent in the cases that have been studied and lead to the same effective potential. Indeed, one would expect the first- and second-quantized effective theories to be equivalent as long as quantum effects such as particle creation are discarded. We shall review the demonstration of equivalence below. 

Note that the scattering amplitude contains, in general, not only classical and quantum terms, as identified by their behavior in terms of $\hbar$, but also super-classical terms. With our conventions, classical terms have a finite limit as $\hbar\to0$ and quantum terms vanish, while super-classical contributions give rise to singular expressions, corresponding to infinitely rapid phase oscillations in the $S$-matrix. It is therefore crucial that the super-classical terms cancel out in the procedure of extracting the classical potential from the scattering amplitude. We find that this cancellation occurs and in fact also ensures that the potential is real.

In this work  we show that indeed both the eikonal exponentiation and the reality of the classical potential are  ultimately direct consequences of the unitarity of the quantum theory.

This observation also lies behind the explanation of the following puzzling question: In the Hamiltonian approach one only needs to compute the classical part of the scattering amplitude up to the given order of the expansion in Newton's coupling constant $G_N$. Classical Hamilton-Jacobi analysis then yields the scattering angle up to that order. Why, then, does the eikonal approach require the computation of the near-forward scattering amplitude to all orders in the coupling $G_N$ in order to derive a fixed-order result for scattering angle? One of the purposes of this paper is to provide an answer to this question. For consistency, it must be that the exponentiation of all higher order terms required in the eikonal limit is automatic. We shall argue that the infinite string of identities needed for the eikonal exponentiation of the classical parts of the near-forward scattering amplitude follow from unitarity. This then resolves the apparent conflict and explains why the two methods for calculating the scattering angle are equivalent.

We consider the scattering problem in a general $D$-dimensional setting rather than just limiting ourselves to the four-dimensional case. As is known already from non-relativistic quantum mechanics,  four space-time dimensions represents a borderline case for scattering in Coulomb-like potentials (such as the  leading-order scattering in general relativity) due to the slow fall-off of the potential at infinity and the associated logarithmic phase of the scattered wave. In relativistic quantum field theory this is reflected in the well-known infrared divergences of the scattering amplitude in four dimensions. Once we move beyond four dimensions, even just  infinitesimally such as in dimensional regularization, these infrared divergences are regularized.

The need to maintain reparametrization (gauge) invariance at all stages of the amplitude calculations while taming the infrared divergences thus leads us to perform the amplitude calculations beyond $D=4$ dimensions. Moreover, as we shall demonstrate, it is not correct that the $D$-dimensional result just trivially mimics the corresponding one in four-dimensional space-time. A new term proportional to $(D-4)$  appears at one-loop order. This could potentially have repercussions at higher loop order if cancelled against infrared sub-divergences, thus threatening to introduce new finite pieces even after taking the $D \to 4$ limit.

To be more specific, we use the relativistic Lippmann--Schwinger equation to derive the long-range effective potential up to 2PM order from the elastic scattering amplitude of two scalar particles with arbitrary masses  in a generic $D$-dimensional space-time. 

While in Ref.~\cite{Paolodue} the box and triangle diagrams were computed for small transferred momentum $q$, \emph{i.e.} in the classical limit, using a saddle-point evaluation in the space of Schwinger parameters, we here perform the same calculation employing the so-called method of regions \cite{Smirnov} in momentum space.
This consists in evaluating the asymptotic expansion of the relevant Feynman integrals as $q\to0$ considering loop momenta $k$ that scale in a definite way with respect to $q$ in this limit.

We identify the soft region, characterized by the scaling relation $k\sim \mathcal O(q)$, as the one producing the non-analytic terms that eventually give rise to the long-range potential, namely the ones considered in  Ref.~\cite{Paolodue}. 
The integrals also receive contributions from the hard region, $k\sim \mathcal O(1)$, that are proportional to positive integer powers of $q^2$ and hence do not contribute to the long-range behavior in position space, although they are needed for the overall consistency of the small-$q$ expansion. Indeed, as is often the case, the hard and soft series separately possess spurious singularities that are just artifacts of the splitting into regions. However, only the singularities originally present in the Feynman integrals survive in the sum of the two series, which provides a nontrivial cross check of the asymptotic expansion thus obtained. 

Another region that is often used in order to extract the non-analytic terms in the classical limit is the potential region. Considering a combination of classical limit $q\to0$ and nonrelativistic limit\footnote{We are grateful to Julio Parra-Martinez and Mikhail Solon for pointing out that the role played by the non-relativistic limit in the definition of the potential region was not properly spelled out in an earlier version of this paper.} $v\to0$,  with $v$ the characteristic velocity of the asymptotic states in the center-of-mass frame, one defines the scaling of the loop momenta $k=(k^0, \vec k)$ in the potential region as $k^0\sim\mathcal O(q v)$ and $\vec k \sim \mathcal O(q)$. The potential expansion allows one to break down the Feynman integrals into $(D-1)$-dimensional integrals in a non-relativistic spirit. In its turn, this opens the possibility to compare General Relativity amplitudes directly to the $(D-1)$-dimensional integrals arising in the effective theory, \emph{i.e.} to perform the matching mentioned above at the level of integrands, disposing with the need to actually evaluate certain integrals. 
We check that, to leading order in the small-$v$ expansion, the result obtained from the potential region agrees with the non-relativistic limit of the one furnished by the soft region. However, we deem more convenient to apply the method of regions in a covariant fashion directly to the $D$-dimensional integrals involved in the evaluation of the fully relativistic amplitude, as outlined above, \emph{i.e.} to base our calculation on the soft and hard regions.

An important new feature that appears in our analysis for $D>4$ is that the 2PM potential receives a nonzero contribution from the sum of the box and crossed box diagrams, which, of course, vanishes if we take $D=4$. This new contribution comes about because of a nontrivial classical term arising from the sum of box and crossed box diagrams that is not exactly compensated by the Born subtraction of the effective theory.
Interestingly, this compensation is exact for any $D$, and thus no new term appears for $D>4$ if we limit ourselves to leading order in the non-relativistic expansion, \emph{i.e.} to the leading term of the potential region.  

Similarly, when we solve the energy equation for the kinematical relation between momentum and position on the classical trajectory, $p^2(r,G_N)$, in dimensions $D>4$, we find that new terms that are quadratic in the scattering amplitude appear. 
To 2PM order, this nonlinearity is precisely canceled by a new term for the classical scattering angle. In this somewhat surprising way, the scattering angle still depends linearly on the amplitude, to this order. 
The scattering angle we compute here coincides perfectly with the one obtained in Ref.~\cite{Paolodue} using instead the eikonal method.

The paper is organized as follows. 
In Sect.~\ref{sect1} we collect the classical and super-classical terms to the one-loop two-to-two amplitude, arising from triangle and box diagrams, which we evaluate with the method of regions.
In Sect.~\ref{LippmannSchwinger} we extract the long-range classical potential at 2PM order from the amplitude solving the Lippmann--Schwinger equation by means of Born subtractions and  describe the equivalence between this technique and the strategy of EFT matching. 
Sect.~\ref{classical} is then devoted to evaluating, given the 2PM potential, the relation $p^2 (r, G_N)$ for the classical trajectory, which we then use in Sect.~\ref{ScatteringAngle} to determine the deflection angle to 2PM order. 
In Sect.~\ref{sec:EikonalLimit} we furnish explicit expressions for the scattering angle given by the 1PM interaction potential for space-time dimensions ranging from four up to ten. 
The paper contains two appendices. In Appendix \ref{app:normalization} we detail our conventions for the normalization of various scattering amplitudes appearing throughout the paper, while in Appendix \ref{app:regions} we present the explicit calculation of the relevant one-loop integrals in the limit $\hbar \rightarrow 0$ using the method of the regions.

\section{Scattering amplitudes in $D$-dimensional General Relativity}
\label{sect1}

In this section we derive the super-classical and classical parts of the one-loop amplitude   
$\mathcal{M}_{\mathrm{1-loop}}$ in Einstein gravity minimally coupled to two massive scalar fields,
\begin{equation}\label{key}
S = \int d^Dx\,\frac{\sqrt{-g}\,R}{16\pi G_N} - \frac{1}{2} \int d^Dx	\,\sqrt{-g} \sum_{i=1,2}\left( g^{\mu\nu} \hbar^2 \partial_\mu \phi_i \partial_\nu \phi_i + m_i^2 \phi_i^2 \right)\,,
\end{equation}
for a general space-time dimension $D$. Focusing on the gravitational interaction of spin-less fields we can compute the large-distance classical scattering
of Schwarzschild black holes (or more generically a point-particle) in the perturbative loop expansion. This amplitude has been recently computed  in Ref.~\cite{Paolodue} using a Schwinger parametrization of the various propagators entering the loop and the method of steepest descent in those parameters. One of the surprising results was that the classical piece of $\mathcal{M}_{\mathrm{1-loop}}$ includes, for $D > 4$,  a nonvanishing contribution  from the sum of box and crossed-box Feynman diagrams. We here employ an alternative method that, in the QCD literature, is known as the method of the regions~\cite{Smirnov}. It is conveniently used to determine the behavior of a loop integral when one is interested in a kinematical limit involving the external momenta, for instance when one of them is small. Here this method is used to determine an expansion of the loop integrals in powers of $\hbar$, confirming the result of Ref.~\cite{Paolodue}.

Let us consider the scattering of two point-like scalar particles, schematically represented by the diagram in the following figure, whose amplitude is given by a sum over all loop contributions: 
\\
\vspace{5pt}
\begin{fmffile}{MyDiagram}
\begin{align}
\begin{gathered}
\begin{fmfgraph*}(50, 35)
\fmfleft{i1,i2}
\fmfright{o1,o2}
\fmfv{label={$p_1$}, label.dist=1mm}{i1}
\fmfv{label={$p_2$}, label.dist=1mm}{i2}
\fmfv{label={$p_3$}, label.dist=1mm}{o1}
\fmfv{label={$p_4$}, label.dist=1mm}{o2}
\fmf{fermion}{i1,v1}
\fmf{fermion}{i2,v1}
\fmf{fermion}{v1,o1}
\fmf{fermion}{v1,o2}
\fmfblob{.35w}{v1}
\end{fmfgraph*}
\end{gathered}
\Longrightarrow\mathcal{M}(\vec{p},\vec{p}\,')=\sum_{n=0}^{\infty}\mathcal{M}_{n\mathrm{-loop}}(\vec{p},\vec{p}\,') \,.
\end{align}
\end{fmffile}
\vspace{5pt}
\\
We refer to Appendix \ref{app:normalization} for more details on our conventions for the normalization of the scattering amplitude.

In the center-of-mass frame we have
\begin{align}
\begin{matrix*}[l]
p^{\mu}_{1}=(E_{1}(p),\vec{p}\,)\,,\\
p^{\mu}_{2}=(E_{2}(p),-\vec{p}\,)\,,
\end{matrix*}
\qquad
\begin{matrix*}[l]
p^{\mu}_{3}=(E_{1}(p),\vec{p}\,')\,,\\
p^{\mu}_{4}=(E_{2}(p),-\vec{p}\,')
\end{matrix*}
\end{align}
and we define
\begin{equation} \label{para}
p \equiv |\vec{p}\,|=|\vec{p}\,'| \,,
\end{equation}
\begin{equation}
E_{1}(p) \equiv \sqrt{p^{2}+m_{1}^2} \,,\qquad E_{2}(p) \equiv \sqrt{p^{2}+m_{2}^2}\,,
\end{equation} 
\begin{equation}
E_{p} \equiv E_{1}(p)+E_{2}(p)  \,, \qquad \xi(p) \equiv \frac{E_1(p) E_2(p)}{E^2_{p}} \,,
\end{equation}
\begin{equation}
q^\mu \equiv p_1^\mu - p_3^\mu\,,\qquad\vec{q}\equiv\vec{p}-\vec{p}\,'\,.
\end{equation}
The previous quantities are related to the Mandelstam variables 
\begin{equation}
s= - (p_1+p_2)^2\,,\qquad t= - (p_1-p_3)^2 = -q^2
\label{mandelstam}
\end{equation}
and 
\begin{eqnarray}\label{sp2Ep}
s = E_p^2\,,
\qquad
p^2 = \frac{(E_p^2 - (m_1+m_2)^2)(E_p^2 - (m_1-m_2)^2)}{4E_p^2}\,.
\label{comf}
\end{eqnarray}  
We will use a mostly positive signature metric, so that in particular 
\beq
q^{\mu} q_{\mu} =q^2=|\vec{q}\,|^2
\eeq 
in the center-of-mass frame, and following Ref. \cite{Paolodue} we define 
\begin{equation}
\kappa^2_D \equiv 8\pi G_N\,, \qquad \gamma(p^2) \equiv 2(p_1 \cdot p_2)^2-\frac{2m_1^2 m_2^2}{D-2} \,.
\label{gamma(p)}
\end{equation}
We first proceed by decomposing the one-loop amplitude in terms of Feynman integrals as follows:
\begin{equation}
\mathcal{M}_{\mathrm{1-loop}} = d_{\Box} (I_{\Box,s}+I_{\Box,u})+(d_{\triangleleft})_{\mu\nu}I^{\mu\nu}_{\triangleleft}+d_{\triangleleft}I_{\triangleleft}+(d_{\triangleright})_{\mu\nu}I^{\mu\nu}_{\triangleright}+d_{\triangleright}I_{\triangleright}+\cdots\ ,
\label{M1loop}
\end{equation}
where the ellipsis denote quantum contributions. The integrals involved in the above expression are the triangle integrals~\footnote{The dependence on  $\hbar$ in the various integrals follows from the fact that, with our conventions, the amplitude in \eqref{M1loop} has dimension of $E^3 L^{D-1}$ where $E$ is an energy and $L$ is a length, as detailed in Appendix \ref{app:normalization}.} 
\begin{align}
\label{scalartriangle}
I_{\triangleright} &= \int \frac{d^Dk}{(2\pi \hbar)^D}\,\frac{\hbar^5}{(k^2-i\epsilon)\left((q-k)^2-i\epsilon)\right)(k^2-2p_1\cdot k-i \epsilon)}\,,
\\ 
\label{tensortriangle}
I_{\triangleright}^{\mu \nu} &= \int \frac{d^Dk}{(2\pi \hbar)^D}\,\frac{\hbar^3 k^\mu k^\nu}{(k^2-i\epsilon)\left((q-k)^2-i\epsilon)\right)(k^2-2p_1\cdot k-i \epsilon)}\,,
\end{align}
together with $I_{\triangleleft}$, $I_{\triangleleft}^{\mu\nu}$ which are given by substituting $p_1\leftrightarrow p_2$ and $p_3\leftrightarrow p_4$ in Eqs.~\eqref{scalartriangle} and \eqref{tensortriangle}, the box
integral
\begin{equation}
\label{scalarbox}
I_{\Box,s}= \int \frac{d^Dk}{(2\pi \hbar)^{D}}\,\frac{\hbar^5}{(k^2-i\epsilon)((k-q)^2-i\epsilon)(k^2-2p_1\cdot k-i\epsilon)(k^2+2p_2\cdot k-i\epsilon)}
\end{equation}
\newline
and the crossed box $I_{{\Box},u}$, obtained by the replacement $p_1\leftrightarrow -p_3$ from Eq.~\eqref{scalarbox}. The associated decomposition coefficients are 
\beq
 d_{\Box}=4i\kappa_D^4\gamma^2(p^2)\,,\qquad
 d_{\triangleright}^{\mu\nu}=\frac{16i\kappa_D^4(D-3)m_1^4}{(D-2)}\,\frac{\hbar^2p_2^\mu p_2^\nu}{q^2}
\eeq
and
\beq
d_{\triangleright} = 4i m_1^2 \kappa_D^4 \left[2 m_{1}^{2} m_{2}^{2}\, \frac{D^{2}-4 D+2}{(D-2)^{2}}-2 m_{1}^{2} E_p^2+m_{1}^{4}+\left(m_{2}^{2}-E_p^2\right)^{2}\right]\,,
\eeq
while $d_{\triangleleft}^{\mu\nu}$ and $d_{\triangleleft} $ are obtained by replacing $m_1\leftrightarrow m_2$ in $d_{\triangleright}^{\mu\nu}$ and $d_{\triangleright}$.

In Appendix \ref{app:regions} we employ the method of expansion by regions to evaluate the classical limit of the one-loop integrals \eqref{scalartriangle}, \eqref{tensortriangle} and \eqref{scalarbox} in arbitrary dimensions $D$ and in a generic reference frame. This limit entails letting $\hbar\to0$ in such a way that in the center-of-mass frame the three-momentum transfer $\vec{q}$ vanishes, while the transferred wave number $\frac{1}{\hbar}\, |\vec{q}\,|$, the total energy $E_p$ and the masses $m_1$, $m_2$ are kept fixed (see \emph{e.g.} \cite{Kosower, Berndue}). It turns out that this analysis in $D$ dimensions presents some new features
as compared with that of Ref. \cite{Cheung}, while being in perfect agreement for $D=4$. The modified expressions for generic $D \ge 4$ will be instrumental 
in reproducing the correct scattering angle in $D$ dimensions \cite{Paolodue}.

Quoting first for completeness the tree-level contribution 
\begin{equation}
\label{IR}
\mathcal{M}_{\mathrm{tree}}(\vec p, \vec p\,')=-{2\gamma(p^2) \kappa^2_{D}}\,\frac{\hbar^2}{q^2}\,,
\end{equation}
we are finally able to cast the classical and super-classical terms of the one-loop scattering amplitude in General Relativity and in $D$ dimensions in the following form:
\begin{equation}
\mathcal{M}_{\mathrm{1-loop}}(\vec p,\vec p\,')=\mathcal{M}_{\triangleleft}(\vec p,\vec p\,')+\mathcal{M}_{\triangleright}(\vec p,\vec p\,')+\mathcal{M}_{\Box,s}(\vec p,\vec p\,')+\mathcal{M}_{\Box,u}(\vec p,\vec p\,')+\cdots\ ,
\end{equation}
where
\beq
\label{IR1looptr}
\begin{split}
&\mathcal{M}_{\triangleleft}(\vec p, \vec p\,')+\mathcal{M}_{\triangleright}(\vec p, \vec p\,') =- \frac{2\sqrt{\pi} \kappa^4_D(m_1+m_2)}{(4 \pi)^{\frac{D}{2}}} 
\\ 
&\times \left(4(p_1\cdot p_2)^2-\frac{4m_1^2m_2^2}{(D-2)^2}-\frac{(D-3)E_p^2 p^2}{(D-2)^2} \right) \frac{\Gamma(\frac{5-D}{2})\Gamma^2(\frac{D-3}{2})}{\Gamma(D-3)} \left(\frac{q^2}{\hbar^2}\right)^{\frac{D-5}{2}}
\end{split}\eeq
and
\beq
\begin{split}
\mathcal{M}_{\Box, s}(\vec p, \vec p\,')+\mathcal{M}_{\Box, u}(\vec p, \vec p\,')
&=-\frac{i \pi}{(4\pi)^{\frac{D}{2}}}\frac{2\kappa^4_{D}\gamma^2(p^2)}{E_{p} \, p}\frac{\Gamma(\frac{6-D}{2})\Gamma^2(\frac{D-4}{2})}{\Gamma(D-4)}\,\frac{1}{\hbar} \left(\frac{q^2}{\hbar^2}\right)^{\frac{D-6}{2}}
\\
&-\frac{2\sqrt{\pi} \kappa^4_D \gamma^2(p^2)}{(4 \pi)^{\frac{D}{2}}}\frac{(m_1+m_2)}{E^2_{p} \, p^2}\frac{\Gamma(\frac{5-D}{2})\Gamma^2(\frac{D-3}{2})}{\Gamma(D-4)} \left(\frac{q^2}{\hbar^2}\right)^{\frac{D-5}{2}}\,.
\label{IR1loopbox}
\end{split}
\eeq
These results are in agreement with those of Ref. \cite{Paolodue}~\footnote{Actually the corresponding amplitudes in Ref.~\cite{Paolodue} are obtained from the ones appearing here multiplying by a factor $-\frac{1}{\hbar}$, since in this paper we use \eqref{AppA1}, while in Ref.~\cite{Paolodue}  \eqref{alternativenormaliz} is used instead.}. 

It should be stressed that the above result for the triangle and box contributions \eqref{IR1looptr}, \eqref{IR1loopbox} is obtained from the expansion of the corresponding integrals in the \emph{soft} region, as detailed in Appendix \ref{app:regions}. 
Such integrals also receive additional contributions from the \emph{hard} region that are, however, proportional to positive integer powers of $\frac{q^2}{\hbar^2}$. We thus discard such terms because they would give rise to strictly local contributions in position space, while we are interested in the long-range behavior of the effective potential. Nevertheless, the interplay between the soft and the hard series is important because it ensures the proper cancellation of spurious divergences that arise for specific dimensions in the above expressions, \emph{e.g.} in $D=5$, and thus provides a nontrivial consistency check of the asymptotic expansion.

The expression \eqref{IR1looptr} for the triangle topologies could be also alternatively obtained from the leading-order expansion of the associated triangle integrals in the \emph{potential} region, as described in Appendix \ref{PotReg}. 
The potential region also allows for a quick evaluation of the sum of box and crossed box diagrams to leading order in the nonrelativistic limit, $\frac{p}{m_1}, \frac{p}{m_2}\ll1$. 

	The result furnished by the leading potential region coincides with the small-velocity limit of \eqref{IR1loopbox}, which, as we stressed, is based on the  soft region. Actually, the first term on the right-hand side of \eqref{IR1loopbox}, namely the super-classical term, coincides with the corresponding term arising from the leading potential approximation. The second term, instead, agrees with the corresponding classical term in the leading potential expansion only in the nonrelativistic limit, in which $E_p\approx m_1+m_2$.  We refer the reader to  Appendix \ref{PotReg} for a detailed discussion of this comparison.

\section{The Post-Minkowskian potential in arbitrary dimensions}
\label{LippmannSchwinger}

In this section, 
we address the calculation of the long-range effective interaction potential to 2PM order in arbitrary dimension, stressing in particular the new elements that appear when away from $D=4$.  Our strategy is based on the method of Born subtractions \cite{Cristofoliuno, Cristofolidue}, which is equivalent to the technique of EFT matching \cite{Cheung, Berndue}. 

As we have stressed, the two-to-two amplitude presents, to one-loop order, both super-classical, $\mathcal O(\hbar^{-1})$, and classical, $\mathcal O(\hbar^{0})$, contributions, as identified by their $\hbar$ scaling. The super-classical term arises in particular from the sum of box and crossed box diagrams, which are also the source of the imaginary part of the amplitude and, in $D=4$, of the infrared divergence. Inverse powers of $\hbar$ are 
conventionally labelled ``IR'' in four dimensions since  they characterize the terms responsible for infrared divergences there.  It should be stressed, however, that the very notion
of infrared divergent integrals becomes ambiguous away from four dimensions. Therefore, we shall keep labelling the terms entirely by
their scaling (power) with respect to $\hbar$, which is well-defined for any $D$. 

The calculation of the post-Minkowskian potential in the center-of-mass frame 
will then reveal how the super-classical and imaginary term eventually cancel, providing a well-defined, real and classical expression for the interaction potential, but leave behind nontrivial contributions in generic dimensions $D>4$. We will also see how this cancellation can be understood as a consequence of the unitarity of the underlying quantum theory.

\subsection{The Lippmann--Schwinger equation in $D$ dimensions}
\label{solving}

In order to define a post-Minkowskian potential in momentum space and in an arbitrary number of dimensions $D=d+1$ we can use a fully relativistic Lippmann-Schwinger equation as in \cite{Cristofoliuno}
\begin{equation}
\label{lippmann}
\mathcal{{\tilde{M}}}(\vec p,\vec p\,')=\tilde{V}^{D}(\vec p,\vec p \,')+\int\frac{d^d\vec k}{(2\pi\hbar)^d}\frac{\tilde{V}^{D}(\vec p,\vec k) \mathcal{{\tilde{M}}}(\vec k,\vec p\,')}{E_p-E_k + i\epsilon}\,.
\end{equation} 
where in the left-hand side we have defined scattering amplitudes with a proper normalization factor
(see Appendix \ref{app:normalization}, in particular Eq.~\eqref{reducedM})
\beq\label{NRNorm}
\mathcal{\tilde{M}}(\vec p,\vec p\,') = \frac{\mathcal{M}(\vec p,\vec p\,')}{4E_{1}(p)E_{2}(p)} \,\, ,
\eeq 
while on the right hand side we have denoted by $ \mathcal{{\tilde{M}}}(\vec k,\vec p\,')$ their analogue definition off the energy shell with $|\vec{k}| \neq |\vec{p}\,'|$.
In what follows our aim is to extract  the classical potential to 2PM order for arbitrary $D\ge4$. We will work in the center-of-mass frame using an isotropic gauge which identifies the phase space $(r,p)$ of a two body Hamiltonian with the Fourier analogue of the exchanged momentum $q$ in the center of mass and with the modulus of the momenta  $p$. The advantage of the latter is the absence of $p \cdot r$ terms in the Hamiltonian and it has shown extremely useful in the computation of post-Minkwoskian Hamiltonians as shown in \cite{Cheung, Cristofoliuno}.\\
We solve perturbatively Eq.~\eqref{lippmann} for the potential itself
\begin{equation}
\label{all}
\begin{split}
&\tilde{V}^{D}(\vec p,\vec p\,')=\mathcal{{\tilde{M}}}(\vec p,\vec p\,')\\
&+\sum_{n=1}^{\infty} (-1)^n\int \frac{d^d \vec k_1}{(2\pi \hbar)^d} \, \frac{d^d \vec k_2}{(2\pi \hbar)^d}  \cdots  \frac{d^d \vec k_n}{(2\pi \hbar)^d} 
\frac{\mathcal{\tilde{M}}(\vec p,\vec k_1) \cdots \mathcal{\tilde{M}}(\vec k_{n},\vec p\,')}{(E_{p}-E_{k_1}+i \epsilon) \cdots(E_{k_{n-1}}-E_{k_{n}}+i \epsilon)}
\end{split}
\end{equation}
and truncate the series up to order $G^2_{N}$
\begin{equation}
\label{potential}
\tilde{V}_{1\mathrm{PM}}^{D}(\vec p,\vec p\,')+
\tilde{V}_{2\mathrm{PM}}^{D}(\vec p,\vec p\,')=\mathcal{\tilde{M}}_{\mathrm{tree}}(\vec p,\vec p\,')+\mathcal{\tilde{M}}_{\mathrm{1-loop}}(\vec p,\vec p\,')+\mathcal{\tilde{M}}_{B}(\vec p,\vec p\,')\,,
\end{equation}
where we have denoted the first Born subtraction by 
\begin{equation}
\label{Born}
\mathcal{\tilde{M}}_{B}(\vec p,\vec p\,') ~\equiv~ -\int\frac{d^d\vec k}{(2\pi\hbar)^d}\frac{\mathcal{\tilde{M}}_{\mathrm{tree}}(\vec p,\vec k)
\mathcal{\tilde{M}}_{\mathrm{tree}}(\vec k,\vec p\,')}{E_p-E_k + i\epsilon}\,.
\end{equation}
Although we do not explicitly distinguish between on-shell and off-shell scattering amplitudes in our notation, it should be stressed that the functions $\mathcal{\tilde{M}}(p,k)$ entering the integrand on the right-hand side of \eqref{Born} are evaluated for states that do not necessarily respect energy conservation and the sum over states indeed runs over all intermediate $(D-1)$-momenta $\vec{k}$. They are defined by $T$-matrix elements for asymptotic states with energies unconstrained, $i.e.$, $|\vec{p}\,| \neq |\vec{k}|$. This is analogous to the EFT approach where the potential $\tilde{V}^{D}(\vec p,\vec k)$  
likewise is defined off the energy shell, $i.e.$, with $|\vec{p}\,| \neq |\vec{k}|$. The off-shell extension of the $T$-matrix and $V$ corresponds to the choice of 
operator basis in the EFT formalism. For instance, insisting on $(D-1)$-dimensional rotational symmetry, the analog of Wilson coefficients in the expansion of $V$ will not 
depend on the scalar product $\vec{p}\cdot\vec{k}$ but only on $\vec{p}^2$ and $\vec{k}^2$. After Fourier transforming, this corresponds to the choice of isotropic coordinates.
In the center-of-mass frame and using this isotropic parametrization
\begin{equation}
\label{offshell}
\mathcal{\tilde{M}}_{\mathrm{tree}}(\vec k,\vec k\,') \equiv G_{N} \frac{A_{1}\left(\frac{k^2+k'^2}{2} \right)}{\frac{1}{\hbar^2}|\vec{k}-\vec{k}'|^2}\,,\qquad{A_1 \bigg(\frac{k^2 + k'^2}{2}\bigg) = - \frac{4 \pi \gamma( \frac{k^2 + k'^2}{2})}{E_1 (\frac{k^2 + k'^2}{2}) E_2 (\frac{k^2 + k'^2}{2} )}}\,,
\end{equation}
where $|\vec{k}|$ is not necessarily equal to $|\vec{k}'|$. For a physical on-shell process in which $|\vec{p}|=|\vec{p}\,'|$ this of course reduces to
\begin{equation}
\label{onshell}
\mathcal{\tilde{M}}_{\mathrm{tree}}(\vec p,\vec p\,') = G_{N} \frac{A_{1}(p^2)}{\frac{1}{\hbar^2}q^2}\,, \qquad A_{1}(p^2) = -\frac{4 \pi \gamma(p^2)}{E_{1}(p)E_{2}(p)}\,.
\end{equation}
At this point we need to evaluate the Born subtraction given by the integral 
in Eq.~\eqref{Born}. 
We focus on the contributions to \eqref{Born} arising from the soft region, which are obtained in this case expanding the integrand around $k^2=p^2$. Indeed, to more directly compare with the discussion of the expansion by regions presented in Appendix \ref{app:regions}, we could let $\vec k = \vec p + \vec \ell$ and then expand for $\vec \ell\sim\mathcal O(\hbar)$, which implies $k^2=p^2+\mathcal O(\hbar)$. One can also check that performing the expansion with respect to this shifted variable $\vec \ell$ eventually leads to the same final answer for the leading and subleading terms. 
We thus begin by Taylor-expanding the denominator and discard quantum terms. In doing so, we find 
\begin{equation}\label{int1lint2l}
\begin{split}
\mathcal{\tilde{M}}_{B}(\vec{p},\vec{p}\, ')
&=-2E_{p}\xi(p) \int\frac{d^d\vec k}{(2\pi\hbar)^d}\frac{\mathcal{\tilde{M}}_{\mathrm{tree}}(\vec p,\vec k)\mathcal{\tilde{M}}_{\mathrm{tree}}(\vec k,\vec p\,')}{ \vec p\,^{2}- \vec k\,^2+i \epsilon}\\
&+ \bigg(\frac{1-3\xi(p)}{2E_{p} \xi(p)} \bigg)\int\frac{d^d\vec k}{(2\pi\hbar )^d}\,\mathcal{\tilde{M}}_{\mathrm{tree}}(\vec p,\vec k)\mathcal{\tilde{M}}_{\mathrm{tree}}(\vec k,\vec p\,')+\cdots\ ,
\end{split}
\end{equation}
where ellipsis denotes quantum contributions which we discard. Using Eq.~\eqref{offshell}, we find
\begin{equation}
\begin{split}
\mathcal{\tilde{M}}_{B}(\vec{p}, \vec{p}\,')
&= 
-2 E_{p} \xi(p)G_{N}^{2}\int\frac{d^dk}{(2\pi\hbar)^d}\frac{ \hbar^4 A^2_{1}\left(\frac{\vec p\,^2+\vec k\,^2}{2}\right)}{(\vec p\,^2-\vec k^2 +i \epsilon)|\vec{k}-\vec{p}\,|^2|\vec{k}-\vec{p}\,'|^2}
\\
&+G^2_{N}\bigg(\frac{1-3\xi(p)}{2E_p \xi(p)} \bigg)
\int\frac{d^dk}{(2\pi\hbar)^d}\frac{ \hbar^4 A^2_{1}\left(\frac{\vec p\,^2+\vec k^2}{2}\right)}{|\vec{k}-\vec{p}\,|^2|\vec{k}-\vec{p}\,'|^2}+\cdots\,.
\end{split}
\end{equation}
We now Taylor-expand also the numerator around $k^2=p^2$. Using Eq.~\eqref{offshell} and reinstating $\kappa_{D}$, we find
\begin{equation}
\label{impo}
\begin{split}
&\mathcal{\tilde{M}}_{B}(\vec p,\vec p\,') =- \frac{\gamma^{2}(p^2) \kappa^4_D}{2E_{p}^3 \xi(p)} \int
\frac{d^dk}{(2\pi\hbar)^d}\frac{ \hbar^4}{(\vec p\,^2-\vec k\,^2+i \epsilon)|\vec{k}-\vec{p}\,|^2|\vec{k}-\vec{p}\,'|^2}
\\
&+\frac{\kappa^4_D}{4E_{p}^3\xi^2(p)} \bigg(\frac{\gamma^{2}(p^2)(\xi(p)-1)}{2E_{p}^2 \xi(p)} - 4\gamma(p) \: 
p_1\cdot p_2 \bigg)\int \frac{d^dk}{(2\pi \hbar)^d}\frac{\hbar^4}{|\vec{k}-\vec{p}\,|^2|\vec{k}-\vec{p}\,'|^2}+\cdots\ ,
\end{split}
\end{equation}
where we have used the following relation,
$
\frac{\partial}{\partial p^2} \gamma (p^2) =  - \frac{2p_1\cdot p_2}{\xi(p)}
$.

The first integral in Eq. (\ref{impo}) is given in Eq.~\eqref{integrale}, while the second can be evaluated with Feynman parameters. The super-classical and classical parts of the Born subtraction to this order can then be written as follows
\beq
\begin{aligned}
\label{Bornsub}
&\mathcal{\tilde{M}}_{B}(\vec p,\vec p\,')= \frac{i \pi \gamma^{2}(p^2) \kappa^4_D 
}{2  p \:\xi(p) E_{p}^3 (4\pi)^{\frac{D}{2}} }\frac{\Gamma\left(\tfrac{6-D}{2}\right)\Gamma^2(\frac{D-4}{2})}{\Gamma(D-4)} \,\frac{1}{\hbar}\left(\frac{q^2}{\hbar^2}\right)^{\frac{D-6}{2}}
\\
&+\frac{\kappa^4_{D}\gamma^{2}(p^2)}{4E_{p}^3p^2\xi(p) (4\pi)^{\frac{D-1}{2}}} \frac{\Gamma\left(\tfrac{5-D}{2}\right)\Gamma^2(\frac{D-3}{2})}{\Gamma(D-4)} \left(\frac{q^2}{\hbar^2}\right)^{\frac{D-5}{2}}\\
&+\frac{\kappa^4_D}{4E_{p}^3\xi^2(p) (4\pi)^{\frac{D-1}{2}}}\bigg(\frac{\gamma^{2}(p^2) (\xi(p)-1)}{2E_{p}^2 \xi(p)}-4 p_1\cdot p_{2} \gamma (p^2) \bigg)\frac{\Gamma^2(\frac{D-3}{2})\Gamma(\frac{5-D}{2})}{\Gamma(D-3)}\left(\frac{q^2}{\hbar^2}\right)^{\frac{D-5}{2}}+\cdots
\end{aligned}
\eeq
where again ellipsis denotes quantum contributions. Remarkably, not only do  the box and crossed box diagrams give nonvanishing super-classical and classical contributions for $D \neq 4$, but similar contributions are also contained in the Born subtraction. It turns out, as expected, that the two super-classical contributions exactly cancel each other. The classical terms, however, remain and reproduce for $D=4$  the result of {Ref.~\cite{Cristofoliuno}}. 

The cancellation of the (super-classical) imaginary part can be interpreted as a consequence of unitarity. Indeed, applying the relation \eqref{unitarity-constraint-T} to the two-to-two scattering in the center-of-mass frame, one has
\beq
\mathcal{\tilde M}(\vec p,\vec p\,')-\overline{\mathcal{\tilde M}(\vec p\,',\vec p\,)}=-i2\pi \int \frac{d^{d}\vec k}{(2\pi\hbar)^{d}}{\delta(E_p-E_k)}\overline{\mathcal{\tilde M}(\vec k,\vec p\,)}  \mathcal{\tilde M}(\vec k,\vec p\,')\,.
\eeq 
Recalling that the tree-level amplitude is real and that, because of time reversal invariance,  the whole invariant amplitude is symmetric under the exchange of $\vec p$ and $\vec p\,'$, we then have, to 2PM order,
\beq\label{Imagine2to2}
\mathrm{Im} \, \mathcal{\tilde M}_{\mathrm{1-loop}}(\vec p,\vec p\,')
= -\pi \int \frac{d^d \vec k}{(2\pi\hbar)^d} {\delta(E_p-E_k)}\,\mathcal{\tilde M}_{\mathrm{tree}}(\vec p, \vec k\,)  \mathcal{\tilde M}_{\mathrm{tree}}(\vec k,\vec p\,')\,.
\eeq
Comparing the right-hand sides of \eqref{potential} and \eqref{Born}, this identity guarantees that the imaginary part of $\mathcal {\tilde M}_{\mathrm{1-loop}}$ must cancel against that of the Born subtraction $\mathcal {\tilde M}_{B}$.

In conclusion, we get the following potential in momentum space up to 2PM:
\begin{equation}
\label{Potential}
\begin{split}
&{\tilde{V}}_{1\mathrm{PM}}^{D} (\vec{p}, \vec{p}\,')+ {\tilde{V}}_{2\mathrm{PM}}^{D} (\vec{p}, \vec{p}\,' )= - \frac{\gamma (p^2) \kappa_D^2 \hbar^2}{2 \xi(p) E_p^2 q^2} \\
+&\frac{\kappa_D^4 (m_1+m_2)}{(4\pi)^{\frac{D-1}{2}} 4 \xi(p) E_p^2}  \left(\!\!-4(p_1\cdot p_2)^2\!  +\frac{4m_1^2 m_2^2}{(D-2)^2} +\frac{(D-3) E_p^2 p^2}{(D-2)^2}\right)\frac{\Gamma (\frac{5-D}{2}) \Gamma^2 (\frac{D-3}{2})}{\Gamma (D-3)} \left( \frac{q^2}{\hbar^2}\right)^{\frac{D-5}{2}} \\
+&\frac{\kappa_D^4 }{4E_p^3 \xi^2(p) (4\pi)^{\frac{D-1}{2}}} \left( \frac{\gamma^2(p^2) ( \xi(p) -1)}{2 E_p^2 \xi(p)} - 4 p_1\cdot p_2 \gamma (p^2)\right)\frac{\Gamma (\frac{5-D}{2}) \Gamma^2 (\frac{D-3}{2})}{\Gamma (D-3)} \left( \frac{q^2}{\hbar^2}\right)^{\frac{D-5}{2}} \\
-&\frac{\kappa_D^4 \gamma^2 (p^2) (m_1 +m_2 - E_p)}{(4\pi)^{\frac{D-1}{2}}\xi(p) E_p^4 p^2}  
\frac{\Gamma (\frac{5-D}{2}) \Gamma^2 (\frac{D-3}{2})}{\Gamma (D-4)} \left( \frac{q^2}{\hbar^2}\right)^{\frac{D-5}{2}}\,. 
\end{split}
\end{equation}
Fourier-transforming it to  configuration space,
\begin{equation}
V^{D}(r,p) = \int \frac{d^{d}\vec q}{(2\pi\hbar)^d}\tilde{V}^{D}(\vec p,\vec p\,') e^{\frac{i}{\hbar}\,\vec q\cdot \vec x} \,,
\end{equation} 
and making use of the identity
\begin{equation}\label{FThom}
\int \frac{d^{d} \vec q}{(2\pi\hbar)^{d}}\left(\frac{q^2}{\hbar^2}\right)^{\nu} e^{\frac{i}{\hbar}\,\vec q\cdot \vec x}=\frac{2^{2 \nu}}{\pi^{\frac{d}{2}}}\frac{\Gamma(\nu+\frac{d}{2})}{\Gamma(-\nu)}\frac{1}{r^{2\nu+d}} ~,
\end{equation} 
we get the potential in configuration space up to order 2PM
\begin{align}
V_{}^{D}(r,p)&=V_{1\mathrm{PM}}^{D}(r,p)+V_{2\mathrm{PM}}^{D}(r,p)+\cdots \ ,
\\
\label{treedia}
V_{1\mathrm{PM}}^{D}(r,p)&=-\frac{\gamma(p^2) G_N}{E_{p}^2 \xi(p)}\frac{\Gamma(\frac{D-3}{2})}{\pi^{\frac{D-3}{2}}}\frac{1}{r^{D-3}}\,, 
\end{align}
\begin{equation}
\label{equazioni}
\begin{split}
&V_{2\mathrm{PM}}^{D}(r,p)\\
&=\frac{G^2_N(m_1+m_2)}{\pi^{D-3}E_{p}^2 \xi(p)}
\bigg(\frac{4m^2_1m^2_2}{(D-2)^2}+\frac{(D-3)[(p_1\cdot p_2)^2-m_1^2m_2^2]}{(D-2)^2}-4(p_1\cdot p_2)^2 \bigg) \frac{\Gamma^2(\frac{D-3}{2})}{r^{2D-6}}\\
&+\frac{G_N^2}{E_{p}^3\xi^2(p)}\bigg(\frac{\gamma^{2}(p^2) (\xi(p)-1)}{2E_{p}^2 \xi(p)}-4\gamma(p^2)
p_1 \cdot p_{2} \bigg)\frac{\Gamma^2(\frac{D-3}{2})}{ \pi^{D-3}}\frac{1}{r^{2D-6}}\\
&+\frac{G_N^2\gamma^{2}(p^2) (E_{p}-m_{1}-m_{2}) }{E_{p}^4\: p^2\: \xi(p)\pi^{D-3}} \frac{\Gamma^2(\frac{D-3}{2})}{\Gamma(D-4)}\frac{\Gamma(D-3)}{r^{2D-6}}\,.
\end{split}
\end{equation}
Let us stress once more that, for $D > 4$,  the 2PM potential thus receives a nontrivial contribution from box and crossed-box diagrams that is not exactly compensated by the Born subtraction. The combination of the two is proportional to the difference between the total energy and the sum of the masses as shown in the last line of Eq.~\eqref{equazioni}. 
As we shall see in the next section, the appearance of this term for $D > 4$ will give rise to a modification in the linear relation between the classical part of the amplitude 
and the expression for $p^2(r, G_N)$ in the classical trajectory that exists in $D = 4$ dimensions \cite{Kalin, Cristofolidue}. 

\subsection{The Effective Field Theory Matching in $D$ dimensions}

In the previous section we have shown how the classical effective potential can be obtained from a scattering amplitude by means of the Born subtraction, which involves inverting \eqref{lippmann} perturbatively. We have seen in particular how the potential acquires new nontrivial terms at 2PM order in higher dimensions. Let us now briefly explain how this calculation can be performed following the method of EFT amplitude-matching introduced in \cite{Cheung}.

We consider two theories: a fundamental one, which we also call the underlying theory, of two massive scalar fields minimally coupled to gravity, and an effective theory of two massive scalars interacting through a four-point interaction potential, which we denote by $\tilde{V}^{D}(\vec p,\vec p\,')$ in momentum space.

In this approach, one starts by making an ansatz for the effective potential: to 2PM order and in momentum space one has
\beq\label{potTOmatch}
\tilde{V}^{D}(\vec p, \vec p\,')= 
G_N c_1\!\left(\tfrac{p^2+p'^2}{2}\right) \left(\frac{q^2}{\hbar^2}\right)^{-1}+ G_N^2 c_2\!\left(\tfrac{p^2+p'^2}{2}\right) \left(\frac{q^2}{\hbar^2}\right)^{\frac{D-5}{2}}+\cdots\ ,
\eeq
where $c_1$ and $c_2$ are unknown coefficients. Since the fundamental and the effective theory  should give rise to the same {dynamics} for the massive scalar particles, a valid matching condition between the two is the equality of scattering amplitudes order by order in the coupling, or equivalently in the PM counting
\begin{equation}
\label{Mtomatch}
\mathcal{\tilde{M}}_{(n-1)\mathrm{-loop}}(\vec p,\vec p\,')=\mathcal{M}_{n\mathrm{PM}}^{\mathrm{EFT}}(\vec p,\vec p\,') \,,   
\end{equation}
where the left hand side of Eq.~\eqref{Mtomatch} is computed in the full theory with the normalization of Eq.(\ref{NRNorm}), while the right hand side is computed in the effective theory by a perturbative expansion in iterated bubbles  as done in \cite{Cheung}.  
At 1PM order, comparing the coefficient of $G_N$ in \eqref{Mtomatch} with the tree amplitude \eqref{IR}, as dictated by the matching condition
\beq
\mathcal{\tilde{M}}_{\mathrm{tree}}(\vec p, \vec p\,')=\mathcal{M}^{\mathrm{EFT}}_{1\mathrm{PM}}(\vec p, \vec p\,')\,,
\eeq
gives 
\beq\label{matchuno}
c_1(p^2)=A_1(p^2)
\eeq
with $A_1(p^2)$ as in \eqref{onshell}. 

At 2PM order, the EFT amplitude is made by two contributions, a contact term proportional to the potential and a bubble: truncating at $G_{N}^{2}$ order one finds
\beq
\begin{split}
\label{2pmeft}
\mathcal M^{\mathrm{EFT}}_{2\mathrm{PM}}(\vec p, \vec p\,')
&= G_N^2 c_2\!\left(p^2\right) \left(\frac{q^2}{\hbar^2}\right)^{\frac{D-5}{2}}\\
&+G_N^2\int \frac{d^d\vec k}{(2\pi\hbar)^d} \frac{\hbar^4 c^2_1\!\left(\tfrac{p^2+k^2}{2}\right)}{(E_p-E_k+i\epsilon)|\vec p - \vec k|^2 |\vec p\,'-\vec k|^2}+\cdots\ ,
\end{split}\eeq
At this point one needs to evaluate the integral appearing in the second line of \eqref{2pmeft} and then compare this the EFT amplitude with the box and triangle contributions \eqref{IR1looptr}, \eqref{IR1loopbox} so as to derive $c_2(p^2)$. 
However, thanks to the condition \eqref{matchuno}, the second line of \eqref{2pmeft} equals $-\widetilde{\mathcal M}_{B}(\vec p, \vec p\,')$, namely the Born subtraction \eqref{Born} except for the overall sign. Therefore the matching condition
\begin{equation}
 {\mathcal{\tilde{M}}}_{\mathrm{1-loop}}(\vec p, \vec p\,')={\mathcal{M}}^{\mathrm{EFT}}_{2\mathrm{PM}} (\vec p, \vec p\,') 
\end{equation}
is equivalent to
\begin{equation}
\label{levelmatch}
\mathcal{\tilde{M}}_{\triangleleft}(\vec p,\vec p\,')+\mathcal{\tilde{M}}_{\triangleright}(\vec p,\vec p\,')+\mathcal{\tilde{M}}_{\Box,s}(\vec p,\vec p\,')+\mathcal{\tilde{M}}_{\Box,u}(\vec p,\vec p\,')=\tilde{V}_{2\mathrm{PM}}^{D}(\vec p,\vec p\,')
-\widetilde{\mathcal M}_{B}(\vec p, \vec p\,')\,.
\end{equation}
We thus see that the EFT matching condition is in fact identical to Eq.~\eqref{potential}, which was at the basis of the calculation of the previous subsection, and thus leads to the same answer for the 2PM potential \eqref{Potential}.

Let us once again briefly stress the new features arising in this analysis in higher dimensions.
We find that the box topologies not only provide a super-classical term that is compensated by a corresponding contribution in the effective theory, but also possess a subleading term which is non vanishing and classical in $D>4$. This term is not removed by a similar contribution from $\mathcal{M}_{B}(\vec{p},\vec{p}\,')$ and this leaves a term in the 2PM potential which is proportional to the difference in the total energy and masses. This term vanishes at $D=4$, as can be seen from the last line of Eq.~\eqref{equazioni}.

\subsection{More on the EFT matching and the Lippmann-Schwinger equation}
At $2$PM and in arbitrary dimensions the classical post-Minkwoskian potential describing a binary system in isotropic coordinates is equivalent if computed using the Lippmann-Schwinger equation or the EFT matching. Restricting to the conservative sector, we can easily show the equivalence to hold to all orders in $G_{N}$ and in arbitrary dimensions. To this extent, let's go back to Eq.~\eqref{lippmann} and let's find a formal solution for a given scattering amplitude $\mathcal{M}(\vec{p},\vec{p}\,')$. Similar to Eq.~\eqref{all}, the potential will be given by a formal series 
\begin{equation}
\label{formalsolution}
\begin{split}
&\mathcal{{\tilde{M}}}(\vec p,\vec p\,')=\tilde{V}^{D}(\vec p,\vec p\,')\\
&+\sum_{n=1}^{\infty}\int \frac{d^d \vec k_1}{(2\pi \hbar)^d} \, \frac{d^d \vec k_2}{(2\pi \hbar)^d}  \cdots  \frac{d^d \vec k_n}{(2\pi \hbar)^d} 
\frac{\tilde{V}^{D}(\vec p,\vec k_1) \cdots \tilde{V}^{D}(\vec k_{n},\vec p\,')}{(E_{p}-E_{k_1}+i \epsilon) \cdots(E_{k_{n-1}}-E_{k_{n}}+i \epsilon)}\,.
\end{split}
\end{equation}
At this point, we can recast each propagator in Eq.~\eqref{formalsolution} as being an ``effective two body propagator'' so as to rewrite each of them as a couple of matter propagators 
\begin{equation}
\label{decprop}
\frac{1}{E_{k_i}-E_{k_{j}}}=i \int \frac{dk_{0}}{2 \pi} \frac{1}{k_0-\sqrt{k^2_{j}+m^2_1}}\frac{1}{E_{k_i}-k_0-\sqrt{k^2_{j}+m^2_2}}\,.
\end{equation}
If we now plug back Eq.~\eqref{decprop} into Eq.~\eqref{formalsolution} we can easily recognize on the right hand side of the latter the same scattering amplitude computed in \cite{Cheung}, where the $n^{\mathrm{th}}$ term of the series corresponds to the $n^{\mathrm{th}}$ loop in an effective field theory of only scalar fields. Using this observation, we get
\begin{equation}
\mathcal{{\tilde{M}}}(\vec p,\vec p\,')=\mathcal{{\tilde{M}}}^{EFT}(\vec p,\vec p\,')
\end{equation} 
thus showing the equivalence between EFT matching and the Lippmann-Schwinger equation. It would be interesting to understand if the equivalence persists once introducing radiative effects in the potential, which are expected to first appear at 4PM \cite{Bernuno}.
\section{From the classical amplitude to kinematics}
\label{classical}
In the previous section we have used the classical limit of the scattering amplitude to derive the classical potential at 2PM order. Including the kinetic terms this brings us to the following Hamiltonian describing the interaction between the two objects with mass $m_1$ and $m_2$:
\begin{equation}
H (r,p) = \sum_{i=1,2} \sqrt{p^2 + m_i^2} +V^{D}_{1\mathrm{PM}}(r,p) + V^{D}_{2\mathrm{PM}}(r,p)=E\,. 
\label{energy}
\end{equation}
Since $E$ is a constant of motion the previous equation implicitly determines the quantity $p^2 = p^2 (r,G_{N})$ as a function of $r$ and $G_N$. Knowledge of this function is crucial in order to compute the scattering angle $\chi$ in the center-of-mass frame. Going to polar coordinates
we can write $p^2$ as follows:
\begin{equation}
p^2(r,G_{N})=p^2_{r}+\frac{L^2}{r^2} \,,
\label{polaco}
\end{equation}
where $p_\phi \equiv L$ is the conserved angular momentum of the system.
Then, the deflection angle is given by the relation:
\begin{equation}
\chi=-2\int_{r_{\mathrm{min}}}^{+\infty}\frac{\partial p_{r}}{\partial L}\,dr -\pi = 2L \int_{r_{\mathrm{min}}}^\infty \frac{dr}{r^2 p_r} - \pi\,,
\label{defle}
\end{equation}
$r_{\mathrm{min}}$ being the positive root of $p_{r}$ closest to zero. 
As noticed in Refs.~\cite{Bernuno,Berndue,Kalin,Cristofolidue,Damourtre}  for $D=4$ one has the remarkable relation 
\begin{equation}
p^2(r,G_N)=p^2_{\infty}-2 E_{p_{\infty}} \xi(p_{\infty})\tilde{\mathcal{M}}(r,p_{\infty})\,,
\label{Kinematics}
\end{equation}
where $\tilde{\mathcal{M}}(r,p_{\infty})$ is the Fourier transform of the amplitude given by
\begin{equation}
\tilde{\mathcal{M}}^{\mathrm{cl.}}(r,p) ~\equiv~ \int \frac{d^d \vec q}{(2\pi \hbar)^d}\mathcal{\tilde{M}}^{\mathrm{cl.}}(\vec p,\vec p\,')e^{i\frac{\vec q}{\hbar}\cdot \vec x} \,.
\label{Fouriertra}
\end{equation}

Working as usual in the center-of-mass frame, we find it convenient here to emphasize the difference between the momentum evaluated along the classical trajectory, $p^2(r,G_N)$, and the asymptotic momentum by the denoting the latter by $p_\infty$, although it had been simply called $p$ in Sect.~\ref{sect1}. For instance, the relation \eqref{sp2Ep} between the asymptotic momentum and the energy now reads
\begin{equation}
p^2_{\infty} = \frac{(m_1^2+m_2^2-E^{2}_{p_{\infty}})^2-4m_{1}^{2}m_{2}^{2}}{4E^{2}_{p_{\infty}}} \,.
\end{equation}

We shall now generalize  Eq. \eqref{Kinematics}  to the $D$-dimensional case. 
Starting from Eq.~\eqref{energy}, we expand the function $p^2(r,G_N)$, whose existence is ensured by the implicit function theorem, order by order in the coupling $G_N$. This allows us to write
\begin{equation}
\label{kine1}
p^2(r,G_{N})=p^2_{\infty}+G_N\ (p^2)'_{G_N=0}(r)+\frac{G_N^2}{2}\ (p^2)''_{G_N=0}(r)+\cdots\ ,
\end{equation} 
where for brevity 
	\begin{equation}
(p^2)'_{G_N=0}(r)=\frac{\partial}{\partial G_N}p^2(r,G_N)\big|_{G_N=0}\,,\qquad \frac{1}{2}(p^2)''_{G_N=0}(r)=\frac{1}{2}\frac{\partial^2}{\partial G_N^2}p^2(r,G_N)\big|_{G_N=0}
	\end{equation}
denote the first two coefficients of said expansion in powers of $G_N$.
Note that \eqref{kine1} is a $D$-independent expression. We then extend the analysis of Ref.~\cite{Cristofolidue}, substituting \eqref{kine1} in \eqref{energy} and solving order by order in $G_N$, to get
\begin{equation}
\label{1}
G_{N}(p^2)'_{G_N=0}(r)=-2 E_{p_{\infty}} \xi(p_{\infty}) V^{D}_{1\mathrm{PM}}(r,p) \big|_{p^2=p^2_{\infty}}
\end{equation}
and
\begin{equation}\label{double}
\begin{split}
&\frac{G^2_{N}}{2}(p^2)''_{G_N=0}(r)=-2E_{p_{\infty}}\xi(p_{\infty}) \bigg[V_{2\mathrm{PM}}^D(r,p)\\
&-2E_{p} \xi(p) V_{1\mathrm{PM}}^D(r,p)\partial_{p^2}
V_{1\mathrm{PM}}^D(r,p)+\bigg(\frac{3\xi(p)-1}{2E_{p}\xi(p)} \bigg)(V^D_{1\mathrm{PM}})^2(r,p)\bigg]_{p^2=p^2_{\infty}}\,.
\end{split}
\end{equation}
Using {the fact that $\gamma (p^2)$ in Eq.~\eqref{gamma(p)} can be written as follows,
\begin{equation}
\gamma (p^2) = 2 E_p^2 p^2 + 2 m_1^2 m_2^2 \frac{D-3}{D-2}\,,
\label{gamp2}
\end{equation}
we can easily get
\begin{eqnarray}
\partial_{p^2} \left( \frac{\gamma (p^2)}{E_1 (p) E_2 (p)}\right) = - \frac{\gamma (p^2) (1-2\xi(p))}{2\xi^3(p) E^4_{p}} + \frac{2}{\xi (p)} \left(1+ \frac{p^2}{\xi(p)E^2_p} \right)\,.
\label{partp2}
\end{eqnarray}
Inserting then in Eqs.~\eqref{1}  and \eqref{double}   the potential in Eq.~\eqref{equazioni}, }
we find:
\begin{equation}
G_{N}(p^2)'_{G_N=0}=-2 E_{p_{\infty}} \xi(p_{\infty})\:\bigg[ -\frac{\gamma(p_{\infty}^2) G_N}{E^2_{p_{\infty}} \xi(p_{\infty})}\frac{\Gamma(\frac{D-3}{2})}{\pi^{\frac{D-3}{2}}}\frac{1}{r^{D-3}}\bigg]= -2 E_{p_{\infty}} \xi(p_{\infty}) \tilde{\mathcal{M}}^{\mathrm{cl.}}_{\mathrm{tree}}(r,p_{\infty})
\label{G1}
\end{equation}
together with
\begin{equation}\label{G2}
\begin{aligned}
&\frac{G^2_{N}}{2}(p^2)''_{G_N=0}
=- 2E_{p_{\infty}}\xi(p_{\infty}) 
\bigg[-\frac{G^2_N}{\pi^{D-3}}\frac{\Gamma^2(\frac{D-3}{2})}{r^{2D-6}}\frac{(m_1+m_2)}{E^2_{p} \xi(p)}\bigg(4(p_1\cdot p_2)^2-\frac{4m^2_1m^2_2}{(D-2)^2}\\
&-\frac{(D-3)E_p^2 p^2}{(D-2)^2} \bigg)+\frac{G_N^2\gamma^{2}(p^2) (E_{p}-m_1-m_2) }{E^4_{p}\: p^2 \xi(p)\: \pi^{D-3}} \frac{\Gamma^2(\frac{D-3}{2})}{\Gamma(D-4)} \frac{\Gamma(D-3)}{r^{2D-6}} 
\bigg]_{p=p_{\infty}}\\
&=-2E_{p_{\infty}} \xi(p_{\infty})  \bigg( \tilde{\mathcal{M}^{\mathrm{cl.}}}_{\triangleleft, \triangleright}(r,p_{\infty})+(\tilde{\mathcal{M}}^{\mathrm{cl.}}_{\mathrm{tree}})^2(r,p_{\infty})\frac{\xi(p_{\infty})(E_{p_{\infty}}-m_1-m_2)}{p_{\infty}^2}\frac{\Gamma(D-3)}{\Gamma(D-4)} \bigg)\\
&=-2E_{p_{\infty}} \xi(p_{\infty})  \bigg( \tilde{\mathcal{M}}^{\mathrm{cl.}}_{\mathrm{1-loop}}(r,p_{\infty})+(\tilde{\mathcal{M}}^{\mathrm{cl.}}_{\mathrm{tree}})^2(r,p_{\infty})\frac{\xi(p_{\infty}) E_{p_{\infty}}}{p_{\infty}^2}\frac{\Gamma(D-3)}{\Gamma(D-4)}  \bigg)\,,
\end{aligned}
\end{equation}
where the Fourier transform of the classical part of the scattering amplitude is defined by Eq.~\eqref{Fouriertra}.
Inserting Eqs. \eqref{G1} and \eqref{G2} in Eq.~\eqref{kine1}, we get
\begin{multline}
\label{NewDamour}
p^2(r,G_{N})=p^2_{\infty}-2E_{p_{\infty}} \xi(p_{\infty}) \bigg(\tilde{\mathcal{M}}^{\mathrm{cl.}}_{\mathrm{tree}}(r,p_{\infty})+\tilde{\mathcal{M}}^{\mathrm{cl.}}_{\mathrm{1-loop}}(r,p_{\infty})\\+(\tilde{\mathcal{M}}^{\mathrm{cl.}}_{\mathrm{tree}})^2(r,p_{\infty})\frac{\xi(p_{\infty}) E_{p_{\infty}}}{p_{\infty}^2}\frac{\Gamma(D-3)}{\Gamma(D-4)} \bigg) +\cdots\ ,
\end{multline}
which of course reduces to Eq.~\eqref{Kinematics} for $D=4$.

It was argued in Ref. \cite{Cristofolidue} that the simpler relation in four dimensions nicely aligned with our expectations that the effective potential describing the scattering of particles from flat space at minus infinity to flat space at plus infinity should depend only on the classical part of the scattering amplitude. We note that this expectation, although slightly modified due to the new term proportional to the square of the tree-level amplitude at 2PM order, is still borne out by this new result for $D > 4$.

\subsection{An alternative derivation}

An alternative derivation of the modified relation \eqref{NewDamour} for $D > 4$ that directly points towards a
generalization to any order in the post-Minkowskian expansion proceeds via Damour's effective Hamiltonian defined by the solution to the energy equation \eqref{energy}~\cite{Damouruno,Damourdue}.

To apply this strategy, let us  start with the following ansatz $p^2(r,G_{N})$ for the solution of Eq.~\eqref{energy}
\begin{equation}
\label{Damoursformula}
p^2(r,G_{N})=p^2_{\infty}+\sum_{n=1}^{2}\frac{G^{n}_{N}f_{n}^{D}(p^2_{\infty})}{r^{n(D-3)}} \,,
\end{equation}
where the constants $f_{n}^{D}$ are found by solving Eq.~\eqref{energy} iteratively. 
As discussed in detail in Refs.~\cite{Damourdue,Kalin,Cristofolidue}, one can consider the energy-momentum relation \eqref{Damoursformula} as an effective
nonrelativistic ``Hamiltonian'' for the scattering problem, in which the term $p^2_\infty$ is regarded as the kinetic term, \emph{i.e.} the unperturbed Hamiltonian, while  
\begin{equation}\label{effpot2}
V_{\mathrm{eff}} \equiv -\sum_{n=1}^{2}\frac{G^{n}_{N}f_{n}^{D}(p^2_{\infty})}{r^{n(D-3)}}
\end{equation}
plays the role of an effective small perturbation.
Notice however that the ``potential'' $V_{\mathrm{eff}}$ has the dimension of an energy squared by \eqref{Damoursformula}.
It is crucial that here the coefficients of the potential are constants, only depending on the total conserved energy $E$.

The associated Lippmann--Schwinger equation then reads
\begin{equation}
\label{lipp}
\mathcal{\tilde{M}}_{\mathrm{eff}}(\vec p,\vec p\,')=\tilde{V}_{\mathrm{eff}}(\vec p,\vec p\,')+\int
\frac{d^d\vec k}{(2\pi\hbar)^d}\frac{\mathcal{\tilde{M}}_{\mathrm{eff}}(\vec p,\vec k\,)\tilde{V}_{\mathrm{eff}}(\vec k,\vec p\,')}{\vec p\,^2-\vec k^2+i \epsilon}\,,
\end{equation} 
where we have rescaled the amplitude by a normalization factor according to 
\beq
\tilde{\mathcal{M}}_{\mathrm{eff}}(r,p_{\infty}) = 2E_{p_{\infty}} \xi(p_{\infty})\tilde{\mathcal{M}}(r,p_{\infty})
\eeq
as in \eqref{reducedMeff} and $\tilde{V}_{\mathrm{eff}}$ denotes the effective potential in momentum space.
In four dimensions the perturbative iteration of Eq.~\eqref{lipp} produces only super-classical terms. For example, at 2PM order, the perturbative expansion of Eq.~\eqref{lipp}
\begin{equation}
\mathcal{\tilde{M}}_{\mathrm{eff}}(\vec p,\vec p\,')=\tilde{V}_{\mathrm{eff}}(\vec p,\vec p\,')+\int
\frac{d^3\vec k}{(2\pi\hbar)^3}\frac{\tilde{V}_{\mathrm{eff}}(\vec p,\vec k\,)\tilde{V}_{\mathrm{eff}}(\vec k,\vec p\,')}{\vec p\,^2-\vec k^2+ i \epsilon}+\cdots
\end{equation}
implies 
\begin{equation}
\label{look}
\mathcal{\tilde{M}}_{\mathrm{eff}}(\vec p,\vec p\,')=\tilde{V}_{\mathrm{eff}}(\vec p,\vec p\,') +
 \int
\frac{d^3\vec k}{(2\pi\hbar)^3}\frac{16\pi^2 (f_{1})^2G_N^2 \hbar^4}{(\vec p\,^2-\vec k^2+ i \epsilon)(\vec{k}-\vec{p}\,)^{2}(\vec{k}-\vec{p}\,')^{2}}+\cdots\ ,
\end{equation}
where $f_1$ stands for $f_1^D$ for $D=4$ and we have used that the Fourier transform of 
$\frac{1}{r}$ is equal to $ \frac{4\pi \hbar^2}{q^2}$ (see Eq.~\eqref{FThom}). From Eq.~\eqref{integrale} one can see that the integral in the previous equation has only super-classical and quantum contributions in $D=4$, or in other words that its classical piece vanishes in four dimensions. 

However, this argument does not apply for arbitrary dimensions $D > 4$. Working again to 2PM order, the integral involved is now
\begin{equation}\label{lookD}
\mathcal{\tilde{M}}_{\mathrm{eff}}(\vec p,\vec p\,')=\tilde{V}_{\mathrm{eff}}(\vec p,\vec p\,') +\frac{1}{\Gamma\left(\tfrac{D-3}{2}\right)^2}\int
\frac{d^d\vec k}{(2\pi\hbar)^d}\frac{16 \pi^{D-1}G_{N}^{2}(f_{1}^D)^2\hbar^4}{(\vec p\,^2-\vec k^2+ i \epsilon)(\vec{k}-\vec{p}\,)^{2}(\vec{k}-\vec{p}\,')^{2}}+\cdots \,,
\end{equation}
where we employed \eqref{FThom}.
Using Eq.~\eqref{integrale} and restricting ourselves to just the classical part of this equation, we get in position space,
\begin{equation}
\label{equo}
\mathcal{\tilde{M}}^{\mathrm{cl.}}_{\mathrm{eff}}(r,p)=V_{\mathrm{eff}}(r,p) -\frac{1}{2p^2}\frac{\Gamma(D-3)}{\Gamma(D-4)}\frac{G_{N}^{2}(f_{1}^D)^2}{r^{2(D-3)}}
\end{equation}
from which
\begin{equation}
\label{read}
V_{\mathrm{eff}}(r,p)=\mathcal{\tilde{M}}^{\mathrm{cl.}}_{\mathrm{eff}}(r,p)+\frac{1}{2p^2}\frac{\Gamma(D-3)}{\Gamma(D-4)}(\mathcal{\tilde{M}}^{\mathrm{cl.}}_{\mathrm{eff}, \mathrm{tree}})^2(r,p)\,.
\end{equation}
Inserting the proportionality relation $\tilde{\mathcal{M}}_{\mathrm{eff}}(r,p_{\infty}) = 2E_{p_{\infty}} \xi(p_{\infty})\tilde{\mathcal{M}}(r,p_{\infty})$, we obtain that the effective potential at 2PM order for $p=p_{\infty}$ is
\begin{equation}
\label{effe}
\begin{aligned}
V_{\mathrm{eff}}(r,p_{\infty}) &\equiv 2E_{p_{\infty}} \xi(p_{\infty})\\ &\bigg(\tilde{\mathcal{M}}^{\mathrm{cl.}}_{\mathrm{tree}}(r,p_{\infty})+\tilde{\mathcal{M}}^{\mathrm{cl.}}_{\mathrm{1-loop}}(r,p_{\infty})+
(\tilde{\mathcal{M}}^{\mathrm{cl.}}_{\mathrm{tree}})^2(r,p_{\infty})\frac{\xi(p_{\infty}) E_{p_{\infty}}}{p_{\infty}^2}\frac{\Gamma(D-3)}{\Gamma(D-4)} \bigg)
\end{aligned}
\end{equation}
as well as the relation 
\begin{equation}\begin{aligned}
p^2(r,G_{N})&=p^2_{\infty}-2E_{p_{\infty}} \xi(p_{\infty}) \bigg(\tilde{\mathcal{M}}^{\mathrm{cl.}}_{\mathrm{tree}}(r,p_{\infty})\\
&+\tilde{\mathcal{M}}^{\mathrm{cl.}}_{\mathrm{1-loop}}(r,p_{\infty})+(\tilde{\mathcal{M}}^{\mathrm{cl.}}_{\mathrm{tree}})^2(r,p_{\infty})\frac{\xi(p_{\infty}) E_{p_{\infty}}}{p_{\infty}^2}\frac{\Gamma(D-3)}{\Gamma(D-4)} \bigg) ~,
\end{aligned}\end{equation}
confirming the previous derivation of Eq.~\eqref{NewDamour}. The advantage of this alternative derivation is that it is more suitable to generalization to higher orders in the PM
expansion. Further corrections of arbitrarily high order in $G_N$ will in general appear in the relation when $D>4$.
\section{The Scattering Angle in arbitrary dimensions}
\label{ScatteringAngle}

In this section we  compute the deflection angle and in particular we see how the new terms that appear in the quantity  $p^2(r,G_{N})$ reproduce the deflection angle  already obtained from the eikonal in dimensions greater than four \cite{Paolodue}.

For the calculation of the scattering angle using $p^2(r,G_{N})$, one could in principle employ Eqs.~\eqref{polaco} and \eqref{defle}, which however involves computing the root $r_{\mathrm{min}}$ of a polynomial in $G_{N}$ of increasing complexity. A more convenient strategy, as seen in \cite{Cristofolidue}, is to express the scattering angle only in terms of $p^2(r,G_{N})$ and the impact parameter $b$ as \footnote{For an alternative way to relate $p^2(r,G_{N})$ to the scattering angle, see Ref.~\cite{Kalin}.} 
\begin{equation}
\chi_{D} = \sum_{k=1}^{\infty}\tilde{\chi}_k(b)\,, 
\qquad 
\tilde{\chi}_{k}(b)=
\frac{2b}{k!}\int_0^{\infty}du\left(\frac{d}{db^2}\right)^k\frac{(V_{\mathrm{eff}}(r,p_{\infty}))^kr^{2(k-1)}}{p_{\infty}^{2k}}\,, 
\label{chiformula}
\end{equation}
where $r^2=u^2+b^2$, while the effective potential is given by
\begin{equation}
V_{\mathrm{eff}}(r,p_{\infty}) ~=~ -\sum_{n=1}^{\infty}\frac{G^{n}_{N}f_{n}^{D}(p^2_{\infty})}{r^{n(D-3)}}\,,
\end{equation}
which avoids the need to evaluate $r_{\mathrm{min}}$.
Since $p^2(r,G_{N})=p^2_{\infty}-V_{\mathrm{eff}}$, one can always read the $f_{n}^{D}$ coefficients from Eq.~\eqref{effe}.\footnote{In certain dimensions particular combinations of $f_{n}^{D}$ terms in the expansion of the scattering angle may vanish \cite{Cristofolidue}. This phenomenon occurs already at 2PM order in four dimensions, where the expansion of the scattering angle exceptionally does not involve $f_1^2$. This is not so in dimensions $D > 4$.} 

At 2PM order the $D$-dimensional scattering angle is thus provided by 
\begin{equation}
\chi^{2\mathrm{PM}}_{D}=\tilde{\chi}_{1}(b)+\tilde{\chi}_{2}(b)\,,
\end{equation}
where
\begin{align}
\label{equation1}
\tilde{\chi}_{1}(b) &=\frac{2b}{p^2_{\infty}}\int_{0}^{+\infty}du \frac{dV_{\mathrm{eff}}}{db^2}(r,p_{\infty})\,,\\
\label{equation2}
\tilde{\chi}_{2}(b) &=\frac{b}{p^4_{\infty}}\int_{0}^{+\infty}du \: \bigg(\frac{d}{db^2} \bigg)^2 \bigg[r^2V_{\mathrm{eff}}^2(r,p_{\infty}) \bigg]\,.
\end{align}
\newline
From Eq.~\eqref{effe} we can read off the $f_{n}^{D}$ coefficients in terms of the amplitudes, namely 
\begin{equation}
\label{tree}
f_1^{D}(p_{\infty})=\frac{2\gamma(p^2_{\infty})}{E_{p_{\infty}} \pi^{\frac{D-3}{2}}}\Gamma\bigg(\frac{D-3}{2}\bigg)
\end{equation}
and
\begin{equation}
\label{loop}
\begin{aligned}
f_{2}^{D}(p_{\infty})&=\frac{2(m_1+m_2)\Gamma^2\left(\tfrac{D-3}{2}\right)}{E_{p_{\infty}} \pi^{D-2}} \bigg(4(p_{1}\cdot p_{2})^2-\frac{4m^2_1m^2_2+(D-3)p^2 E_p^2}{(D-2)^2} \bigg)_{\!\!p=p_\infty}\\
&+\frac{2\gamma^{2}(p_{\infty}) (m_1+m_2-E_{p_{\infty}}) }{E^3p_{\infty}^2\pi^{D-3}} \Gamma^2\bigg(\frac{D-3}{2}\bigg)\frac{\Gamma(D-3)}{\Gamma(D-4)} \,.
\end{aligned}
\end{equation}
\newline
The integrals in Eqs.~\eqref{equation1}--\eqref{equation2} are elementary. The first contribution to the scattering angle gives
\begin{equation}
\label{scattering angle}
\tilde{\chi}_{1}(b)=\frac{G_Nf_1^D(p_{\infty})}{p_{\infty}^2}\frac{\sqrt{\pi}}{b^{D-3}}\frac{\Gamma(\frac{D-2}{2})}{\Gamma(\frac{D-3}{2})}+\frac{G_N^2f_2^D(p_{\infty})}{p^2_{\infty}}\frac{\sqrt{\pi}}{b^{2D-6}}\frac{\Gamma(D-\frac{5}{2})}{\Gamma(D-3)}\,.
\end{equation}
Inserting Eqs.~\eqref{tree}--\eqref{loop}, this becomes 
\begin{equation}\label{chi1corr}
\begin{aligned}
\tilde{\chi}_{1}(b)&=\frac{2\gamma(p_{\infty}) G_N}{p_{\infty}^2\:  E_{p_{\infty}}\: b^{D-3}}\frac{\Gamma(\frac{D}{2})}{\pi^{\frac{D-4}{2}}}\\
&+\frac{2G^2_N\Gamma(D-\frac{5}{2})\Gamma^2(\frac{D-3}{2})}{p_{\infty}^2\: E_{p_{\infty}}\: b^{2D-6}\pi^{D-\frac{7}{2}}}\frac{(m_1+m_2)}{\Gamma(D-3)} \bigg(4(p_{1}\cdot p_{2})^2
-\frac{4m^2_1m^2_2+(D-3)p^2 E_p^2}{(D-2)^2} \bigg)_{\!\!p=p_\infty}
\\
&+\frac{2\gamma^{2}(p_{\infty}) (m_1+m_2-E_{p_{\infty}}) }{E^3_{p_{\infty}}p_{\infty}^4\pi^{D-\frac{7}{2}}} \Gamma^2\bigg(\frac{D-3}{2}\bigg)\frac{\Gamma(D-\frac{5}{2})}{\Gamma(D-4)} \frac{G^2_N}{b^{2D-6}}\,.
\end{aligned}
\end{equation}
The remaining contribution gives
\begin{equation}\begin{aligned}\label{chi2corr}
\tilde{\chi}_{2}(b)&=\frac{bG^2_{N}(f_{1}^{D})^2}{p^4_{\infty}}\int_{0}^{+\infty}du \: \bigg(\frac{d}{db^2} \bigg)^2 r^2\bigg(\frac{1}{r^{2d-4}} \bigg)
\\
&=\frac{2\gamma^{2}(p_{\infty})}{E^2_{p_{\infty}}p^4_{\infty}}\frac{\Gamma(D-\frac{5}{2})}{\Gamma(D-4)}\frac{\Gamma^2(\frac{D-3}{2})}{\pi^{D-\frac{7}{2}}}\frac{G^2_{N}}{b^{2D-6}}\,.
\end{aligned}\end{equation}
Note that this additional term vanishes in four space-time dimensions $D=4$.
Adding these pieces together, we find the $D$-dimensional scattering angle at 2PM order to be 
\begin{equation}
\label{Paoloformula}
\begin{aligned}
\chi^{2\mathrm{PM}}_{D}&=\frac{2\gamma(p_{\infty}) G_N}{p_{\infty}^2\:  E_{p_{\infty}}\: b^{D-3}}\frac{\Gamma(\frac{D}{2})}{\pi^{\frac{D-4}{2}}}
\\
&+\frac{2G^2_N\Gamma(D-\frac{5}{2})\Gamma^2(\frac{D-3}{2})}{p_{\infty}^2\: E_{p_{\infty}}\: b^{2D-6}\pi^{D-\frac{7}{2}}} \frac{(m_1+m_2)}{\Gamma(D-3)} \bigg(4(p_{1}\cdot p_{2})^2-\frac{4m^2_1m^2_2+(D-3)E_p^2 p^2}{(D-2)^2+}\bigg)_{\!\!p=p_\infty}
\\
&+\frac{2\gamma^{2}(p_{\infty}) (m_1+m_2) }{E^3_{p_{\infty}}p_{\infty}^4\pi^{D-\frac{7}{2}}} \Gamma^2\bigg(\frac{D-3}{2}\bigg)\frac{\Gamma(D-\frac{5}{2})}{\Gamma(D-4)} \frac{G^2_N}{b^{2D-6}}
\end{aligned}
\end{equation}
in complete agreement with the eikonal calculation \cite{Paolodue}. 

It is also interesting to see how this agreement comes about. On the one hand, the new classical
pieces from the box and crossed-box diagrams in $D > 4$ dimensions yield a contribution proportional  to $(m_1+m_2 - E_{p_\infty})$ in the last line of Eq.~\eqref{chi1corr}. On the other hand, for $D>4$ there is a new term in the formula for the scattering angle that is proportional to $E_{p_{\infty}}$ (and $f_1^2$) in Eq.~\eqref{chi2corr}. Adding these two contributions one gets the last line of Eq.~\eqref{Paoloformula} where we see that the two terms proportional to $E_{p_{\infty}}$ have cancelled  each other leaving  only the term proportional to $m_1+m_2$.

Finally, let us consider an alternative route to the computation of the scattering angle which also can be phrased in terms of amplitude evaluations and which has been described in Ref.~\cite{Kosower}. As shown there, one can express the change in four-momentum of a particle in two-body scattering by means of
\begin{equation}
\left\langle\Delta p_{1}^{\mu}\right\rangle=\left\langle\psi\left|S^{\dagger} \mathbb{P}_{1}^{\mu} S\right| \psi\right\rangle-\left\langle\psi\left|\mathbb{P}_{1}^{\mu}\right| \psi\right\rangle
\end{equation}
where $S$ denotes the S-matrix and the two particle state is given by a suitable $\ket \psi$.
Re-expressing the S-matrix in terms of the $T$-operator, one gets \cite{Kosower}
\begin{equation}
\label{Deltap}
\left\langle\Delta p_{1}^{\mu}\right\rangle=I_{(1)}^{\mu}+I_{(2)}^{\mu}
\end{equation}
\begin{equation}
\label{idef}
I_{(1)}^{\mu} \equiv  \left\langle\psi\left|i\left[\mathbb{P}_{1}^{\mu}, T\right]\right| \psi\right\rangle \quad , \quad I_{(2)}^{\mu} \equiv \left\langle\psi\left|T^{\dagger}\left[\mathbb{P}_{1}^{\mu}, T\right]\right| \psi\right\rangle
\end{equation}
In the center-of-mass frame, it is now straightforward to relate the scattering angle $\theta$ to Eq.(\ref{Deltap}) by  means of
\begin{equation}\label{defsca}
\sin\theta =\frac{\left\langle\Delta p_{1}^{\mu}\right\rangle b_{\mu}}{p_{\infty} b}
\end{equation}
where $b^{\mu}=(0,\vec{b})$ denotes the impact parameter as in \cite{Kosower}. 
In the case of classical General Relativity and to second Post-Minkwoskian order the scattering angle can be read off from of Eq.~\eqref{defsca} and \eqref{Deltap}, 
\begin{equation}
\theta_{2PM}= \frac{I_{1}^{\mu} b_{\mu}}{p_{\infty} b} +\frac{I_{2}^{\mu} b_{\mu}}{p_{\infty} b}\,,
\end{equation}
since $\sin\theta\simeq \theta$ at this level of approximation.

To this order the scattering angle arises from two contributions, one linear and one quadratic in the involved scattering amplitudes. The term quadratic in the amplitude plays  a role somewhat analogous of the Born subtraction needed to define the potential as in Eq.~\eqref{Born}. Indeed, the quadratic term removes a classically singular term coming from $I^{\mu}_{1}$ \cite{Kosower}, thus rendering a well-defined classical observable in the same way as the Born subtraction of Eq.~\eqref{Born} removes super-classical pieces, and thus allowing the $\hbar \to 0$ limit.  It would be interesting to understand the precise relationship between these two methods, and in particular to see how the method of Ref.~\cite{Kosower} leads to the same result as the two other amplitude methods, also for $D>4$.

\subsection{Eikonal exponentiation and unitarity}
As we have already pointed out in the introduction, the computation of the scattering angle to a certain fixed order in the expansion parameter $G_N$ requires the calculation of an infinite series of terms of the scattering amplitude, in the eikonal approach. This is needed in order to ensure the exponentiation of terms in impact-parameter space. In contrast, the fixed-order calculation that uses the Hamiltonian language needs only the amplitude computed up to the given order in $G_N$. It is therefore instructive to further explore the connection between unitarity, as encoded in Eq.~\eqref{unitarity-constraint-T} and the eikonal exponentiation. 

To analyze this issue, let us consider again the identity \eqref{Imagine2to2} for two-to-two scattering in the center-of-mass frame, which we may recast as
\begin{equation}\begin{aligned}
\label{opt1}
\mathrm{Im} \, \mathcal{M}_{\mathrm{1-loop}}(\vec p,\vec p\,')
&=- \frac{\pi}{2E_{p}}  \int \frac{d^d\vec k}{(2\pi\hbar)^d} \delta(\vec p\,^2-\vec k^2)\mathcal{M}_{\mathrm{tree}}(\vec p,\vec k\,)  \mathcal{M}_{\mathrm{tree}}(\vec k,\vec p\,')\,,
\end{aligned}\end{equation}
(note that we are dealing here with the invariant amplitude $\mathcal M$ instead of $\widetilde{\mathcal M}$) or
\begin{equation}\begin{aligned}
\label{opt1riscr}
\mathrm{Im} \, \mathcal{M}_{\mathrm{1-loop}}(\vec p,\vec p\,')
&=\frac{1}{2E_{p}}  \,\mathrm{Im}\,\int \frac{d^d\vec k}{(2\pi\hbar)^d} \frac{\mathcal{M}_{\mathrm{tree}}(\vec p,\vec k\,)  \mathcal{M}_{\mathrm{tree}}(\vec k,\vec p\,')}{\vec p\,^2-\vec k^2+i\epsilon}\,.
\end{aligned}\end{equation}
The integral appearing on the right-hand side is the same as that in the first line of Eq.~\eqref{int1lint2l}, thus immediately giving us
\begin{equation}
\mathrm{Im} \, \mathcal{M}_{\mathrm{1-loop}}(\vec p,\vec p\,')= \frac{G^2_{N} c_{1}^2(p^2)\pi^{1-\frac{D}{2}}}{2^{D+1} pE_p}\frac{\Gamma\left(\tfrac{6-D}{2}\right)\Gamma^2(\frac{D-4}{2})}{\Gamma(D-4)} \left(\frac{q^2}{\hbar^2}\right)^{\frac{D-6}{2}}\,.
\end{equation}
Transforming to impact parameter space $b$ by means of a Fourier transform in $D-2$ dimensions yields
\begin{equation}
\mathrm{Im}\, \mathcal{{M}}_{\mathrm{1-loop}}(b)=\frac{1}{2}\frac{G^2_{N}c_{1}^2(p^2)}{64 E_p p}\frac{\Gamma^2(\frac{D-4}{2})}{(b^2)^{D-4}} \pi^{2-D}\,,
\end{equation}
while the same Fourier transform for the tree level amplitude \eqref{onshell} gives
\begin{equation}
\mathcal{M}_{\mathrm{tree}}(b)=\frac{G_N c_{1}(p^2)}{4}\frac{\Gamma(\frac{D-4}{2})}{b^{D-4}} \pi^{\frac{2-D}{2}}
\end{equation}
and hence, dividing by the normalization factor $4 E_p p$ as in \cite{Paolodue} (see also Eq.~\eqref{reducedMEik}), we find
\begin{equation}\label{signeik}
\mathrm{Im}\, \frac{\mathcal{{M}}_{\mathrm{1-loop}}(b)}{4 E_p p}=\frac{1}{2}\left( \frac{\mathcal{{M}}_{\mathrm{tree}}(b)}{4E_p p} \right)^2\,.
\end{equation}
This is the first identity needed to ensure exponentiation of the tree-level amplitude in the eikonal limit and we see that it follows from unitarity alone. 

We interpret this as further evidence that, even at higher orders, unitarity indeed lies behind the eikonal exponentiation. A remarkable phenomenon is that in this approach super-classical terms of increasingly high inverse of powers of $\hbar$ are needed to achieve the exponentiation in impact-parameter space that eventually, at the saddle point, leads to the {\em classical} scattering angle.

\section{ {Simple expressions for the deflection angle}}
\label{sec:EikonalLimit}

In this section we show that, if the potential is just given by the contribution of the tree diagram, then we can obtain  a closed expression for the deflection angle in $D$ dimensions.
Let us now assume that the effective potential in $D$ dimensions is only given by the tree-level contribution:
\begin{equation}
\label{treedia1}
V_{\mathrm{eff}}(r)=-\frac{G_N f_1^{D}}{r^{D-3}}\,,
\qquad
 f_1^{D}(p_{\infty})=\frac{2\gamma(p^2_{\infty})}{E_{p_{\infty}} \pi^{\frac{D-3}{2}}}\Gamma\bigg(\frac{D-3}{2}\bigg)\,,
\end{equation}
where $f_1^D$ is given in Eq.~\eqref{tree}.
The deflection angle is computed from Eq.~\eqref{chiformula} which, for the potential in Eq.~\eqref{treedia1}, implies
\begin{equation}\begin{aligned}
\chi^{D}_{\mathrm{tree}}
&=\sum_{k=1}^{\infty}\frac{2b}{k!}\bigg(-\frac{G_Nf_1^D}{p^2_{\infty}}\bigg)^{k}\int_{0}^{+\infty} du\: \partial_{b^2}^{(k)}\bigg[(u^2+b^2)^{k\frac{(5-D)}{2}-1}\bigg] 
\\
&=\sum_{k=1}^{\infty}\frac{2b}{k!}\bigg(\frac{-G_Nf_1^D}{p^2_{\infty}}\bigg)^{k} \:\prod_{l=0}^{k-1}\bigg(k\frac{(5-D)}{2}-1-l \bigg) \int_{0}^{+\infty} du\: \frac{1}{(u^2+b^2)^{1+\frac{k(D-3)}{2}}}\,.
\end{aligned}
\end{equation}
The integral over the variable $u$ can be easily computed and one gets
\begin{equation}
\chi^{D}_{\mathrm{tree}}
=\sum_{k=1}^{\infty}\frac{2b}{k!}\bigg(\frac{-G_Nf_1^D}{p^2_{\infty}}\bigg)^{k} \:\prod_{l=0}^{k-1}\bigg(k\frac{(5-D)}{2}-1-l \bigg) \frac{\sqrt{\pi}}{b^{k(D-3)+1}}\frac{1}{k(D-3)}\frac{\Gamma(\frac{k(D-3)+1}{2})}{\Gamma(\frac{k(D-3)}{2})}\,,
\end{equation}
which we may finally recast in the form
\beq\label{risultato}
\chi^{D}_{\mathrm{tree}}=\sqrt{\pi}\sum_{k=1}^{\infty}\frac{\alpha^k}{k!} \:  \frac{\Gamma(\frac{k(D-3)+1}{2})}{\Gamma(\frac{k(D-5)}{2}+1)}
\end{equation}
with
\beq
\alpha_D=\frac{G_Nf_1^D}{p^2_{\infty}b^{D-3}}\,.
\eeq
In some particular case, such as $D=4,5$, the sum of the series \eqref{risultato} evaluates to simple functions. For $D=4$ one gets \footnote{A closed expression for the scattering angle in $D=4$ up to 2PM included has been given in \cite{Cristofoliuno, Kalin}.}
\begin{equation}
\chi^{4}=2 \arctan\bigg(\frac{\alpha_4}{2}\bigg) \Longrightarrow \tan \frac{\chi^4}{2} = \frac{\alpha_4}{2}\,,
\label{chi4}
\end{equation}
while for $D=5$ one finds
\begin{equation}
\chi^{5}=\frac{\pi}{\sqrt{1-\alpha_5}}-\pi ~.
\label{chi5}
\end{equation}
The two previous deflection angles  have the same form as the deflection angles in Eq.~{(4.5)} of Ref.~\cite{DAppollonio:2010krb} corresponding to the scattering of a massless scalar particle  on a maximally supersymmetric $D6$-brane and on a $D5$-brane, respectively.
For $D=7$ we  get 
\begin{equation}
\chi^{7}=\frac{2 K\left(\frac{4 \sqrt{\alpha_7}}{2 \sqrt{\alpha_7}+1}\right)}{\sqrt{2 \sqrt{\alpha_7}+1}}-\pi\,,
\end{equation}
where $K$ is the complete elliptic integral of first kind. Also this expression agrees with the one in Eq.~{(4.6)} of Ref.~\cite{DAppollonio:2010krb} for the $D3$-brane.
Finally, for $D = 6, 8, 9$ and $D=10$ we can write the deflection angle in terms  of hypergeometric functions:
\begin{align}
\chi^{6} &=2 \alpha_6 \,\:  _3F_2\left(\frac{2}{3},1,\frac{4}{3};\frac{3}{2},\frac{3}{2};\frac{27 \alpha^2_6}{4}\right)+\pi  \, _2F_1\left(\frac{1}{6},\frac{5}{6};1;\frac{27 \alpha_6^2}{4}\right)-\pi\,,
\\
\chi^{8}
&=\pi  \, _4F_3\left(\frac{1}{10},\frac{3}{10},\frac{7}{10},\frac{9}{10};\frac{1}{3},\frac{2}{3},1;\frac{3125 \alpha_8^2}{108}\right) 
\\
&+\frac{8}{3} \alpha_8 \, _5F_4\left(\frac{3}{5},\frac{4}{5},1,\frac{6}{5},\frac{7}{5};\frac{5}{6},\frac{7}{6},\frac{3}{2},\frac{3}{2};\frac{3125 \alpha_8^2}{108}\right)- \pi \,,
\\
\chi^{9}&=\pi\: \:  _2F_1\left(\frac{1}{6},\frac{5}{6};1;\frac{27 \alpha_9}{4}\right)-\pi\,,
\\
\chi^{10} 
&= \pi  \, _6F_5\left(\frac{1}{14},\frac{3}{14},\frac{5}{14},\frac{9}{14},\frac{11}{14},\frac{13}{14};\frac{1}{5},\frac{2}{5},\frac{3}{5},\frac{4}{5},1;\frac{823543 \alpha_{10}^2}{12500}\right)
\\
&+\frac{16}{5} \alpha_{10} \: \:\, _7F_6\left(\frac{4}{7},\frac{5}{7},\frac{6}{7},1,\frac{8}{7},\frac{9}{7},\frac{10}{7};\frac{7}{10},\frac{9}{10},\frac{11}{10},\frac{13}{10},\frac{3}{2},\frac{3}{2};\frac{823543 \alpha_{10}^2}{12500}\right)- \pi\,. 
\end{align}

The power-series expansions of these results (up to order $\alpha^2_D$) again agree with 
Eq.~{(4.8)} of Ref.~\cite{DAppollonio:2010krb} with the following identification of the variables involved in the two cases:
\begin{equation}
\alpha_D \Longleftrightarrow \left( \frac{R_p}{b}\right)^{7-p}\,, \qquad p+D =10\,.
\label{ide}
\end{equation}
An alternative way to show the equivalence between our approach with only the tree diagram potential and that of Ref.~\cite{DAppollonio:2010krb} is using Eq.~\eqref{defle}. In fact in this case $p^2 (r, G_N)$ in Eq.~\eqref{Damoursformula} contains only the term with $n=1$  and taking into account Eq.~\eqref{polaco} one gets the following expression for the deflection angle in Eq.~\eqref{defle}: 
\begin{eqnarray}
\chi_D (b) = 2 \int_{r_{\mathrm{min}}}^\infty \frac{dr}{r^2}  \frac{b}{ \sqrt{ 1 + \left( \frac{R_D}{r}\right)^{D-3} - \frac{b^2}{r^2}}} -\pi
\label{chiD}
\end{eqnarray}
with 
\begin{equation}
b \equiv \frac{L}{ p_\infty}\,, \qquad R_D^{D-3} \equiv  \frac{G_N f_1^D}{p_\infty^2} = \frac{2G_N \gamma (p_\infty^2 )}{E_{p_\infty} p_\infty^2} \frac{\Gamma ( \frac{D-3}{2})}{\pi^{\frac{D-3}{2}}}\,,
\label{bRD}
\end{equation}
where in the last step we have used Eq.~\eqref{treedia1}. 
On the other hand Eq.~{(4.4)} of Ref.~\cite{DAppollonio:2010krb} can be easily rewritten as follows,
\begin{equation}
\chi_p (b) = 2 \int_{r_{\mathrm{min}}}^\infty \frac{dr}{r^2} \frac{b}{ \sqrt{1+ \left(\frac{R_p}{r}\right)^{7-p} - \frac{b^2 }{ r^2}}} - \pi\,,
\label{defleDb}
\end{equation}
where $R_p$ is a quantity defined in Ref.~\cite{DAppollonio:2010krb}.
The two equations give the same deflection angle if we make the following identification:
\begin{equation}
R_p^{7-p} \Longleftrightarrow R_D^{D-3}\,,\qquad
p+D=10\,.
\label{identi}
\end{equation}

\section{Conclusions}
\label{conclusions}

Starting from the elastic scattering amplitude of two scalar particles with arbitrary masses  in Einstein gravity in an arbitrary number $D$ of space-time dimensions, we isolated the terms that contribute in the classical limit by the method of regions. We then extracted from them the long-range classical effective potential between the two scalar particles for arbitrary $D$ by means of the Lippmann--Schwinger equation or, equivalently, by the technique of EFT matching.

We then used the Hamiltonian consisting of the sum of the relativistic kinetic terms for the two particles   and the potential to determine the conjugate momentum $p^2 (r,G_N)$. It turns out that, unlike the case $D=4$, for arbitrary $D$ this relation contains   an extra term proportional to the square of the tree scattering amplitude that, of course, vanishes for $D=4$.  We then used it to compute the deflection angle, finding complete agreement with the one obtained using the eikonal approach~\cite{Paolodue}.  

The approach of this paper is not only different from the one of Ref.~\cite{Paolodue} because here we use the Hamiltonian approach to derive the deflection angle, while Ref.~\cite{Paolodue} was based on the the eikonal approach, but also because the box and crossed box integrals are computed using two different methods. It turns out that, if we use the method of the regions directly on the fully relativistic expression for the box and crossed box diagrams, as explained in Appendix \ref{Box}, we get the  
same result for the subleading term as in Ref.~\cite{Paolodue}, while, if we first go to the potential region  and then compute the subleading term, we get the same result only in the nonrelativistic limit, where the energy of the two particles becomes equal to their mass. Since we use the fully relativistic expression for the sum of the box and crossed box diagrams in the underlying fundamental theory, while the nonrelativistic expression for those diagrams  emerges in the EFT, from the matching between the two theories we get the important result that, for $D > 4$, these diagrams leave 
a nonzero contribution to the potential that, of course, vanishes for $D=4$.
\newline

NOTE ADDED: While this paper was under review a new way to perform the integrals of the potential region appeared \cite{Parra-Martinez:2020dzs}. The authors confirm our $D$-dimensional calculation of the amplitude.

\subsection*{Acknowledgements}

We thank Emil Bjerrum-Bohr, Arnau Koemans Collado, Marios Hadjiantonis, Raffaele Marotta, Julio Parra-Martinez, Rodolfo Russo, Mikhail Solon, Pierre Vanhove, and Gabriele Veneziano for many useful discussions.
This work has received funding from the European Union's Horizon 2020 research and innovation programme under the Marie 
Sklodowska-Curie grant agreement No. 764850 ``SAGEX'' and has also been supported in part by the Danish National Research Foundation (DNRF91). The research of PDV and CH is supported by the Knut and Alice Wallenberg Foundation under grant KAW 2018.0116.

\appendix
\section{Normalization of the amplitude}
\label{app:normalization}

In this Appendix we fix  the conventions that we adopt for the normalization of scattering amplitudes.
We decompose the $S$-matrix according to
\beq
 S= 1 - \frac{i}{\hbar} \, T\,.
\label{AppA1}
\eeq
The operator $T$ has therefore the dimension of an action, $EL$, where $E$ stands for an energy scale and $L$ for a length scale. Its matrix elements $T_{ba}=\langle b| T |a\rangle$ between asymptotic states $|b\rangle$ and $|a\rangle$ define the standard scattering amplitudes $\mathcal M_{ba}$ according to
\beq\label{factorApp}
T_{ba}=(2\pi \hbar)^D\delta(P_{a}-P_{b}) \mathcal M_{ab}\,,
\eeq
where $P_{b}$ and $P_{a}$ denote the total outgoing and incoming $D$-momenta. 
The unitarity of the $S$-matrix $SS^\dagger=1=S^\dagger S$ also implies the following identity among $T$-matrix elements involving the sum over a complete set of intermediate asymptotic states
\beq
T_{ba}-{(T^\dagger)}_{ba} = -\frac{i}{\hbar} \sum_{c}T_{bc}(T^\dagger)_{ca}\,,
\eeq
or, at the level of scattering amplitudes,
\beq\label{unitarity-constraint-T}
\mathcal M_{ab} - \overline{\mathcal M_{ba}} = -i2\pi \sum_{c} (2\pi \hbar)^{D-1} \delta(P_a-P_c)\, \overline{\mathcal M_{ca}}\, \mathcal M_{cb}
\eeq
for states such that $P_a=P_b$.

We are interested in asymptotic states containing two kinds of scalar particles with masses $m_1$ and $m_2$ although we shall suppress the subscripts $1$, $2$ for simplicity. The associated free Hermitian scalar fields $\varphi(x)$ are described by the action
\beq
S_\mathrm{free} = -\frac{1}{2}\int d^Dx \left(\hbar^2 \partial^\mu \varphi\, \partial_\mu \varphi + m^2 \varphi^2\right)\,.
\label{AppA2}
\eeq
The Fock expansion for $\varphi(x)$ can be taken as 
\beq
\label{AppA3}
\varphi(x) = \int\frac{d^{D}p}{(2\pi \hbar)^{D-1}}\,\delta(p^2+m^2) \tilde\varphi(p)e^{\frac{ip\cdot x}{\hbar}}\,,
\eeq
where $\tilde\varphi(p)=a(\vec p\,)$ and $\tilde\varphi(- p)=a^\dagger(\vec p\,)$ for $p=(p^0,\vec p\,)$ and $p^0>0$, while  the canonical commutation relations read 
\beq
\label{commfock}
[a( \vec p),a^\dagger( \vec {p}\,')] =  2E(p) (2\pi\hbar)^{D-1} \delta(\vec p - \vec{p}\,')\,,
\eeq
with $E(p)=\sqrt{{\vec p}\,^{2}+m^2}$ denoting the single-particle energy. The field $\varphi(x)$ has dimension  $E^{-\frac{1}{2}}L^{\frac{1-D}{2}}$ and the creation/annihilation operators $\tilde\varphi(p)$ have dimension $E^\frac{1}{2} L^{\frac{D-1}{2}}$.
Single-particle states are obtained acting with the creation operator $a^\dagger(\vec p\,)$ on the Fock vacuum $|0\rangle$,
\beq
a(\vec p\,)|0\rangle=0\,,\qquad
|\vec p\,\rangle =a^\dagger(\vec p\,)|0\rangle\,,\qquad
\langle \vec p\, |\vec {p}\,'\rangle =  2E(p) (2\pi\hbar)^{D-1} \delta^{(D-1)}(\vec p - \vec {p}\,')\, ,
\label{AppA4}
\eeq
so that their normalization is Lorentz invariant. The completeness relation for asymptotic states reads
\beq
\sum_{n=1}^\infty \int \frac{d^{D-1} \vec p_1}{(2\pi\hbar)^{D-1}} \frac{1}{2E(p_1)}\cdots \frac{d^{D-1} \vec p_n}{(2\pi\hbar)^{D-1}} \frac{1}{2E(p_n)}\,|\vec p_n,\ldots, \vec p_1\rangle \langle \vec p_n,\ldots, \vec p_1|=1\,.
\eeq
The invariant amplitude $\mathcal M( \vec {p}_1,\ldots, \vec {p}_M, \vec p_1\,\!',\ldots, \vec {p}_N\,\!')$ for the scattering of $M$ incoming and $N$ outgoing massive scalars is then given by the relation
\beq
\langle \vec {p}_N\,\!',\ldots, \vec {p}_1\,\!' |  T |\vec p_M,\ldots, \vec p_1\rangle = (2\pi\hbar)^{D}\delta(P-P\,')\mathcal M(  \vec {p}_1,\ldots, \vec {p}_M, \vec p_1\,\!',\ldots, \vec {p}_N\,\!')
\label{AppA5}
\eeq
with
\beq
P=\sum_{i=1}^M p_i\,,\qquad P\,'=\sum_{i=1}^N {p}_i\,\!'
\eeq
and has the physical dimension $EL^{1-D}(EL^{D-1})^{\frac{M+N}{2}}$. This is a direct consequence of the fact that the creation and annihilation operators have dimension $E^\frac{1}{2} L^{\frac{D-1}{2}} $.

For the specific case of two-to-two scattering of particles with mass $m_1$ and $m_2$, which we describe at the beginning of Section \ref{sect1}, one has
\beq\begin{split}
\langle \vec p_4, \vec p_3 | S |\vec p_2, \vec p_1\rangle &= 2E_1(p_1) (2\pi\hbar)^{D-1}\delta(\vec p_1-\vec p_3) 2E_2(p_2) (2\pi\hbar)^{D-1}\delta(\vec p_2-\vec p_4)
\\
&-i2\pi  (2\pi\hbar)^{D-1}\delta( p_1+ p_2- p_3- p_4)\mathcal M(\vec p_1, \vec p_2, \vec p_3, \vec p_4)\,,
\end{split}\eeq
and we adopt a simplified notation for the invariant amplitude evaluated in the center-of-mass frame
\beq
\label{AppA7}
\mathcal M( \vec {p},\vec p\,')= \mathcal M( \vec {p}_1, \vec {p}_2,\vec {p}_3, \vec {p}_4)\,,
\eeq
which has dimension $E^3 L^{D-1}$.
We also consider a reduced $S$-matrix, $s$, which relates to the standard $S$-matrix by
\beq
\langle \vec p_4, \vec p_3 | S |\vec p_2, \vec p_1\rangle = 4E_1(p_1)E_2(p_2) (2\pi\hbar)^{D-1}\delta(\vec p_1+ \vec p_2- \vec p_3- \vec p_4) \langle \vec p\,' | s | \vec p\, \rangle \,,
\eeq
with
\beq
\vec p = \frac{m_2 \vec p_1- m_1 \vec p_2}{m_1+m_2}\,,\qquad
\vec p\,' = \frac{m_2 \vec p_3- m_1 \vec p_4}{m_1+m_2}\,,
\eeq
and reads
\beq\label{reduceds}
\langle \vec p\,' | s | \vec p\, \rangle = (2\pi\hbar)^{D-1}\delta(\vec p-\vec p\,')-i2\pi \delta(E_p-E_{p'}) \widetilde{\mathcal M}( \vec {p}, \vec p\,')
\eeq	
in the center-of-mass frame.
Therefore the reduced amplitude in the center-of-mass frame $\widetilde{\mathcal M}( \vec {p}, \vec p\,')$ is related to the invariant amplitude by
\beq
\label{reducedM}
\widetilde{\mathcal M}( \vec {p}, \vec p\,') = \frac{ \mathcal M(\vec {p}, \vec p\,')}{4E_1( {p})E_2( p)}
\eeq
and has dimension $EL^{D-1}$. Eq.~\eqref{reduceds} for the reduced $S$-matrix can be also written as 
\beq\label{reducedseff}
\langle \vec p\,' | s | \vec p\, \rangle = (2\pi\hbar)^{D-1}\delta(\vec p-\vec p\,')-i2\pi \delta(p^2-p'^2)  \widetilde{\mathcal M}_{\mathrm{eff}}( \vec {p}, \vec p\,')
\eeq
with
\beq\label{reducedMeff}
 \widetilde{\mathcal M}_{\mathrm{eff}}( \vec {p}, \vec p\,') = 2E_{p} \xi(p)\tilde{\mathcal{M}}( \vec {p}, \vec p\,')\,,
\eeq
or as 
\beq\label{reducedsEik}
\langle \vec p\,' | s | \vec p\, \rangle = (2\pi\hbar)^{D-1}\delta(\vec p-\vec p\,')-i2\pi \delta(p-p')  \widetilde{\mathcal M}_{\mathrm{eik}}( \vec {p}, \vec p\,')\,,
\eeq
with
\beq\label{reducedMEik}
{\mathcal M}_{\mathrm{eik}}( \vec {p}, \vec p\,') = \frac{\mathcal M( \vec {p}, \vec p\,')}{4E_p p}\,.
\eeq

We should also mention that the $T$ matrix is often defined in the following \emph{alternative} way:
\beq\label{alternativenormaliz}
S = 1+  i \, T\,.
\eeq
In this case one would get a scattering amplitude that differs from the previous one by a factor $- \hbar$. This alternative normalization was employed in \cite{Paolodue} to retrace the dependence on $\hbar$ of the eikonal factor that one extracts from the scattering amplitude.

\section{One-Loop Integrals in the $\hbar\to0$ Limit}
\label{app:regions}

In this Appendix we explicitly discuss the evaluation of triangle and box integrals in the classical limit $\hbar\to0$, \emph{i.e.} the limit of small transferred momentum $q$. We employ a technique that can be used to extract the asymptotic expansion of Feynman integrals in certain limits known as the method of regions \cite{Smirnov}, which consists in splitting the domain of integration into sectors defined by suitable scaling relations. 

In the examples we shall consider, the asymptotic expansions of Feynman integrals will emerge in particular from the soft region, in which the integrated momentum $k$ scales as $k\sim\mathcal O(q)$, and from the hard region, $k\sim\mathcal O(1)$. The non-analytic contributions in momentum space giving rise to long-range effects in position space, on which we focus in the main body of the paper, are those obtained from the soft region. We will then comment on the relation between the results obtained from these regions and the potential region. This region involves both the classical limit of small $q$ and the nonrelativistic limit of small $v$,  where $v$ is the relative velocity in the center-of-mass frame, and can be characterized by the scaling relations $k^0\sim \mathcal O(qv)$ and $\vec k \sim \mathcal O(q)$.

\subsection{Triangle integrals}
Let us first consider the scalar triangle integral \eqref{scalartriangle}
\beq\label{triangle4regions}
I_{\triangleright} = \int \frac{d^Dk}{(2\pi\hbar)^D}\,\frac{\hbar^5}{(k^2-i\epsilon)\left((q-k)^2-i\epsilon)\right)(k^2-2p_1\cdot k-i\epsilon)}\,,
\eeq
which we may recast as
\beq\label{triangleq4regions}
I_{\triangleright} = \int \frac{d^Dk}{(2\pi\hbar)^D}\,\frac{\hbar^5}{(k^2-i\epsilon)\left((q-k)^2-i\epsilon\right)(k^2-(q_\perp+q)\cdot k-i\epsilon)}
\eeq 
introducing, together with the momentum transfer $q=p_1-p_3$,  the additional variable
\begin{equation}
q_\perp = p_1+p_3\,.
\end{equation} 
Note in particular that $q\cdot q_\perp=0$.

The classical limit consists in letting $\hbar\to0$ in such a way that the momentum transfer $q$ vanishes, while the transferred wave-vector $\frac{1}{\hbar}\, q$ and the average momentum $\frac{1}{2}\,q_\perp$ of the massive particle are kept fixed. We schematically identify this situation by writing
\beq
 q\sim \mathcal O(\hbar)\,,\qquad
 q_\perp \sim \mathcal O(1)\,,\qquad
q\ll q_\perp\,. 
\eeq
We note that this limit requires the mass $m_1$ to be nonzero, in view of the relation
\beq\label{qperpq}
-q_\perp^2 = {4m_1^2}+q^2\,.
\eeq 

We shall now employ the expansion by regions to obtain an asymptotic approximation of the integral \eqref{triangleq4regions} in the classical limit. This method consists in splitting the integration over the loop momentum $k$ into a soft region, characterized by the scaling $k\sim\mathcal O(\hbar)$ and hence $k \sim q \ll q_\perp$, and a hard region, in which $k\sim\mathcal O(1)$ and hence $k \sim q_\perp \gg q$,
namely
\beq
I_\triangleright= I^{(s)}_\triangleright + I^{(h)}_\triangleright,
\eeq
with
\begin{align}
I^{(s)}_{\triangleright} &= \int_{k\sim q} \frac{d^Dk}{(2\pi\hbar)^D}\,\frac{\hbar^5}{(k^2-i\epsilon)\left((q-k)^2-i\epsilon\right)(k^2-(q_\perp+q)\cdot k-i\epsilon)}\,,\\
I^{(h)}_{\triangleright} &= \int_{k\sim q_\perp} \frac{d^Dk}{(2\pi\hbar)^D}\,\frac{\hbar^5}{(k^2-i\epsilon)\left((q-k)^2-i\epsilon\right)(k^2-(q_\perp+q)\cdot k-i\epsilon)}\,.
\end{align} 
One then considers the Taylor expansion of the integrands according to the appropriate scaling relations, thus obtaining two asymptotic series for $I^{(s)}$ and $I^{(h)}$,
\beq\begin{split}
I^{(s)}&=I^{(1s)}+I^{(2s)}+\cdots\ ,\\
I^{(h)}&=I^{(1h)}+I^{(2h)}+\cdots\ .
\end{split}\eeq 
The first two contributions to the soft region thus read
\begin{align}
\label{tr1s}
I_\triangleright^{(1s)} &= \int_{k\sim q} \frac{d^Dk}{(2\pi\hbar)^D}\, \frac{\hbar^5}{(k^2-i\epsilon)((q-k)^2-i\epsilon) (-q_\perp\cdot k-i\epsilon)}\,,\\
\label{tr2s}
I_\triangleright^{(2s)} &= \int_{k\sim q} \frac{d^Dk}{(2\pi\hbar)^D}\, \frac{\hbar^5(-k^2+q\cdot k)}{(k^2-i\epsilon)((q-k)^2-i\epsilon) (-q_\perp\cdot k-i\epsilon)^2}\,,
\end{align}
while for the hard contribution one has
\begin{align}
\label{tr1h}
I_\triangleright^{(1h)} = \int_{k\sim q_\perp} \frac{d^Dk}{(2\pi\hbar)^{D}}\,\frac{\hbar^5}{(k^2-i\epsilon)^2(k^2-q_\perp\cdot k-i\epsilon)}\,,\\
I_\triangleright^{(2h)} = \int_{k\sim q_\perp} \frac{d^Dk}{(2\pi\hbar)^{D}}\,\frac{\hbar^5 q\cdot k(3k^2-2q_\perp\cdot k)}{(k^2-i\epsilon)^3(k^2-q_\perp\cdot k-i\epsilon)^2}\,.
\end{align}
The integration can be then extended to the whole $D$-dimensional space in both regions in view of the fact that the error $R_\triangleright$ thus introduced always takes the form of a scaleless integral and is therefore identically vanishing in dimensional regularization: to leading order, for instance,
\beq
R_\triangleright = \int \frac{d^Dk}{(2\pi\hbar)^{D}} \frac{\hbar^5}{(k^2-i\epsilon)^2(-q_\perp \cdot k-i\epsilon)}=0\,.
\eeq

By means of the above expansion we have reduced the problem to the evaluation of simpler Feynman integrals, which can be directly calculated introducing Feynman parameters and exploiting the orthogonality between $q$ and $q_\perp$, as detailed in Section~\ref{detailsss} below. The leading contribution \eqref{tr1s} to the soft region can be read from the general integral \eqref{formulaperp} and takes the form 
\beq\label{scalarsoft}
I_\triangleright^{(1s)} = \frac{i\sqrt\pi}{m_1(4\pi)^{\frac{D}{2}}}\,\frac{\Gamma\left(\tfrac{D-3}{2}\right)^2\Gamma\left(\tfrac{5-D}{2}\right)}{2\Gamma(D-3)} \left(\frac{q^2}{\hbar^2}\right)^{\frac{D-5}{2}}\,,
\eeq
since $-q_\perp^2=4m_1^2+\mathcal O(\hbar^2)$ thanks to \eqref{qperpq},
while the leading hard contribution \eqref{tr1h} reads, by \eqref{formula12},
\begin{equation}\begin{aligned}\label{IRpole}
I_\triangleright^{(1h)} =\frac{i \Gamma\left(\tfrac{6-D}{2}\right)}{(4-D)(5-D)(4\pi)^{\frac{D}{2}}\hbar}\left(
\frac{m_1^2}{\hbar^2}
\right)^{\frac{D-6}{2}}\,.
\end{aligned}\end{equation}
We note that the leading soft term behaves as $\mathcal O(1)$ as $\hbar\to0$ and is therefore classical, while the hard term scales like $\hbar^{\frac{5-D}{2}}$. Furthermore, the latter is analytic (in fact, constant) in the transferred momentum and therefore corresponds to a local term in position space, while the former gives rise to a power-law dependence on $r$ via \eqref{FThom}. Actually, the whole hard asymptotic expansion is just a power series expansion in $q^2$ and this leads us to focus on the terms arising from the soft region in the discussion of the long-range potential. 

Considering now the subleading soft integral \eqref{tr2s}, we note that the first term in the numerator gives rise to a scaleless integral, after sending $k\to q-k$, and thus can be discarded. The remaining integral is then given by  \eqref{formulaperp2}, namely 
\beq\label{scalarsoftsoft}
I_\triangleright^{(2s)} = -\frac{i\hbar}{m_1^2(4\pi)^{\frac{D}{2}}}\,\frac{\Gamma\left(\tfrac{D-2}{2}\right)^2\Gamma\left(\tfrac{4-D}{2}\right)}{2\Gamma(D-3)} \left(\frac{q^2}{\hbar^2}\right)^{\frac{D-4}{2}}\,,
\eeq
which is $\mathcal O(\hbar)$ and hence quantum. Interestingly, we note that this term of the expansion is divergent as $D\to4$, despite the fact that the original integral \eqref{triangleq4regions} is clearly finite in four dimensions. The appearance of such spurious divergences is a standard feature of the expansion by regions and indicates the presence of cancellations between the soft and the hard series. In this case, the pole at $\epsilon=0$ for $D=4-2\epsilon$ cancels in the sum of the leading hard term \eqref{IRpole} and subleading soft term \eqref{scalarsoftsoft}, leaving behind the finite contribution
\beq\label{cancellation}
\left(I_\triangleright^{(1h)}+I_\triangleright^{(2s)}\right)\big|_{D=4} = \frac{i\hbar}{2m_1^2(4\pi)^{2}}\,\left(
\log\frac{q^2}{m^2_1}-2
\right)\,.
\eeq
This can be regarded as a quantum contribution since it contains terms scaling as $\mathcal O(\hbar\log\hbar)$ and $\mathcal O(\hbar)$ in the classical limit.

A similar strategy also applies to tensor integrals associated to the triangle diagram, such as
\beq
I_\triangleright^{\mu} = 
\int \frac{d^Dk}{(2\pi\hbar)^D}\,\frac{\hbar^{4}k^\mu}{(k^2-i\epsilon)\left[(q-k)^2-i\epsilon\right](k^2-(q_\perp+q)\cdot k-i\epsilon)}
\eeq
and the one appearing in \eqref{tensortriangle},
\beq
I_\triangleright^{\mu\nu} = 
\int \frac{d^Dk}{(2\pi\hbar)^D}\,\frac{\hbar^3 k^\mu k^\nu}{(k^2-i\epsilon)\left[(q-k)^2-i\epsilon\right](k^2-(q_\perp+q)\cdot k-i\epsilon)}\,.
\eeq
After performing a tensor decomposition in terms of $q^\mu$, $q_\perp^\mu$ and $\eta^{\mu\nu}$, these integrals can be evaluated directly in the soft region by means of Feynman parameters, (see \eqref{ellell-q}, \eqref{formulaperp}, \eqref{formulaperp2}, \eqref{formulaperp3}).  To leading order as $\hbar\to0$, one finds
\begin{equation}
\label{vectorsoft}
\begin{split}
I^{(s)\mu}_\triangleright 
&=
\frac{i\sqrt{\pi}}{(4\pi)^{\frac{D}{2}}} \frac{\Gamma\left(\frac{5-D}{2}\right)\Gamma\left(\frac{D-1}{2}\right)\Gamma\left(\frac{D-3}{2}\right)}{2\Gamma(D-2)}
\ \frac{q^\mu}{\hbar m_1} \left(\frac{q^2}{\hbar^2}\right)^{\frac{D-5}{2}}\\
&
+
\frac{i}{(4\pi)^{\frac{D}{2}}} \frac{\Gamma\left(\frac{4-D}{2}\right)\Gamma\left(\frac{D-2}{2}\right)^2}{2\Gamma(D-2)}
\ \frac{p_1^\mu}{m^2_1} \left(\frac{q^2}{\hbar^2}\right)^{\frac{D-4}{2}} 
\end{split}
\end{equation}
and
\begin{equation}\label{tensorsoft}
\begin{aligned}
I_\triangleright^{(s)\mu\nu}&=
\frac{i}{4m_1(4\pi)^{\frac{D}{2}}\Gamma(D-1)}\\
&\times 
\Bigg[
\left(
\eta^{\mu\nu}+\frac{p_1^\mu p_1^\nu}{m_1^2}-(D-1)\frac{q^\mu q^\nu}{q^2}  
\right)
\left(\frac{q^2}{\hbar^2}\right)^{\frac{D-3}{2}}\,{\sqrt\pi\, \Gamma\left(\tfrac{3-D}{2}\right)\Gamma\left(\tfrac{D-1}{2}\right)^2}
\\
&
+\frac{2(q^{\mu}p_1^{\nu}+q^{\nu}p_1^{\mu})}{\hbar m_1}\,\left(\frac{q^2}{\hbar^2}\right)^{\frac{D-4}{2}}\,{\Gamma\left(\tfrac{4-D}{2}\right)\Gamma\left(\tfrac{D-2}{2}\right)\Gamma\left(\tfrac{D}{2}\right)}
\Bigg]\,.
\end{aligned}
\end{equation}
The analogous results for $I_\triangleleft$, $I_\triangleleft^\mu$, $I_{\triangleleft}^{\mu\nu}$ can be obtained by replacing $m_1\leftrightarrow m_2$ in the above expressions \eqref{scalarsoft}, \eqref{scalarsoftsoft}, \eqref{vectorsoft} and \eqref{tensorsoft}.

\subsection{Box integrals}
\label{Box}
Let us now turn to the scalar box integral \eqref{scalarbox}, leaving the $-i\epsilon$ prescription implicit for the time being,
\begin{equation}\label{box4regions}
I_{\Box,s}= \int \frac{d^Dk}{(2\pi\hbar)^{D}}\,\frac{\hbar^5}{k^2(k-q)^2(k^2-2p_1\cdot k)(k^2+2p_2\cdot k)}\,.
\end{equation} 
Introducing the variables
\beq\label{qperpQ}
q_\perp= p_1+p_3\,,\qquad Q = p_1+p_2
\eeq
allows us to recast the desired integral as follows
\beq\label{boxq4regions}
I_{\Box,s}^{(1s)}=
\int \frac{d^Dk}{(2\pi \hbar)^{D}}\,\frac{\hbar^5}{k^2(k-q)^2(k^2+(2Q-q_\perp-q)\cdot k)(k^2-(q_\perp+q)\cdot k)}\,.
\eeq
These new variables satisfy in particular
\beq\label{relqperpQ}
q\cdot q_\perp= 0 =q\cdot Q\,,\qquad
q_\perp\cdot Q = Q^2-(m_1^2-m_2^2)\,.
\eeq

We are interested in the classical limit described by the scaling
\beq
q \sim \mathcal O(\hbar)\,,\qquad  q_\perp,\, Q \sim \mathcal O(1)\,,\qquad q\ll q_\perp,\,Q\,,
\eeq
as $\hbar\to0$, which implicitly requires a nonzero mass because
\beq
-q_\perp^2 = {4m_1^2}+q^2\,.
\eeq
The leading soft term then reads 
\beq\begin{aligned}\label{tobereg}
	I^{(1s)}_{\Box,s} 
	=\int \frac{d^Dk}{(2\pi \hbar)^{D}}\,\frac{\hbar^5}{k^2(k-q)^2((2Q-q_\perp)\cdot k)(-(q_\perp\cdot k))}\,,
\end{aligned}\eeq
where, following the same strategy detailed for the triangle diagram, we have performed a Taylor expansion of the integrand of \eqref{boxq4regions} to leading order for $k\sim\mathcal O(\hbar)$, namely $k\sim q\ll q_\perp,\,Q$.  Introducing a Feynman parameter $x$ for the two linear factors in the denominator, we then have
\beq\begin{aligned}
	I^{(1s)}_{\Box,s} 
	=\int_0^1 dx \int\frac{d^Dk}{(2\pi\hbar)^{D}}\,\frac{\hbar^5}{k^2(k-q)^2((2xQ-q_\perp)\cdot k)^2}\,.
\end{aligned}\eeq
Since $2xQ-q_\perp$ is orthogonal to $q$, we can apply \eqref{formulaperp}, which thus yields
\beq
I^{(1s)}_{\Box,s} 
=
\frac{i\Gamma\left(\tfrac{D-4}{2}\right)^2\Gamma\left(\tfrac{6-D}{2}\right)}{2(4\pi)^{\tfrac{D}{2}}\Gamma(D-4)}\,\frac{1}{\hbar}\left(\frac{q^2}{\hbar^2}\right)^{\frac{D-6}{2}}
\int_0^1\frac{dx}{-\left(xQ-\frac{1}{2}q_\perp\right)^2-i\epsilon}\,,
\eeq
where we have reinstated the $-i\epsilon$ prescription.
The roots of the polynomial 
\begin{equation}
-\left(xQ-\frac{q_\perp}{2}\right)^2-i\epsilon
\end{equation} appearing in the denominator are given up to $\mathcal O(\hbar^2)$ by
\beq\label{roots}
x_\pm = \frac{m_1^2-p_1\cdot p_2 \pm \sqrt{(p_1\cdot p_2)^2 - (m_1m_2)^2}}{m_1^2+m_2^2-2p_1\cdot p_2}
\pm i\epsilon
\eeq
and their real parts both lie in the integration interval, namely between 0 and 1. We thus obtain\footnote{
$\cosh^{-1}(x)=\log(x+\sqrt{x^2-1}).$
}
\beq
I_{\Box,s}^{(1s)} = 
\frac{i\Gamma\left(\tfrac{D-4}{2}\right)^2\Gamma\left(\tfrac{6-D}{2}\right)}{2\hbar(4\pi)^{\tfrac{D}{2}}\Gamma(D-4)}
\ 
\frac{i\pi - \cosh^{-1}\left(-\frac{p_1\cdot p_2}{m_1m_2}\right)}{\sqrt{(p_1\cdot p_2)^2-m_1^2m_2^2}}\,\left(\frac{q^2}{\hbar^2}\right)^{\frac{D-6}{2}}.
\eeq
The crossed box diagram is related to the one we just discussed by $p_1\mapsto -p_3$, which corresponds to exchanging $p_1\cdot p_2 \leftrightarrow -p_1\cdot p_2$ up to $\mathcal O(\hbar^2)$.
The real parts of the roots analogous to \eqref{roots} then no longer fall between $0$ and $1$ and the resulting integral gives
\beq
I_{\Box,u}^{(1s)} = 
\frac{i\Gamma\left(\tfrac{D-4}{2}\right)^2\Gamma\left(\tfrac{6-D}{2}\right)}{2\hbar(4\pi)^{\tfrac{D}{2}}\Gamma(D-4)}
\frac{\cosh^{-1}\left(-\frac{p_1\cdot p_2}{m_1m_2}\right)}{\sqrt{(p_1\cdot p_2)^2-m_1^2m_2^2}}\,\left(\frac{q^2}{\hbar^2}\right)^{\frac{D-6}{2}}.
\eeq
The sum of the leading box and crossed box diagrams finally reads
\beq
\label{Box+CrossedBox}
I_{\Box,s}^{(1s)}+I_{\Box,u}^{(1s)}=\frac{\Gamma\left(\tfrac{D-4}{2}\right)^2\Gamma\left(\tfrac{6-D}{2}\right)}{2\hbar(4\pi)^{\tfrac{D}{2}}\Gamma(D-4)}
\frac{-\pi}{\sqrt{(p_1\cdot p_2)^2-m_1^2m_2^2}}\,\left(\frac{q^2}{\hbar^2}\right)^{\frac{D-6}{2}}.
\eeq

The subleading term in the soft expansion for the box integral is instead
\beq
I_{\Box,s}^{(2s)}
=2\hbar^5\int_0^1 dx
\int \frac{d^Dk}{(2\pi\hbar)^D}\frac{q\cdot k-k^2}{k^2(q-k)^2\left[(2xQ-q_\perp)\cdot k\right]^3}\,,
\eeq
where we have considered the second term in the Taylor expansion of the integrand of \eqref{boxq4regions} for $k\sim\mathcal O(\hbar)$, namely $k\sim q\ll q_\perp,\,Q$.
Recognizing that the second term in the numerator gives rise to a scaleless integral, this expression can be evaluated by the help of formula \eqref{formulaperp2} to
\beq\label{x+x-subbox}
I^{(2s)}_{\Box,s} 
=
-\frac{i\sqrt{\pi}\,\Gamma\left(\frac{5-D}{2}\right)\Gamma\left(\frac{D-3}{2}\right)^2}{4(4\pi)^{\frac{D}{2}}\Gamma(D-4)}\left(\frac{q^2}{\hbar^2}\right)^{\frac{D-5}{2}}\int_0^1\frac{dx}{\left[-\left(xQ-\frac{q_\perp}{2}\right)^2-i\epsilon\right]^\frac{3}{2}}\,.
\eeq
Performing the integral over $x$ then yields, to leading order in $\hbar$,
\beq
I^{(2s)}_{\Box,s} 
=
\frac{i\sqrt{\pi}\,\Gamma\left(\frac{5-D}{2}\right)\Gamma\left(\frac{D-3}{2}\right)^2}{8(4\pi)^{\frac{D}{2}}\Gamma(D-4)}\left(\frac{q^2}{\hbar^2}\right)^{\frac{D-5}{2}}
\frac{\left[
	s\left(\frac{1}{m_1}+\frac{1}{m_2}\right)+(m_1^2-m_2^2)\left(\frac{1}{m_1}-\frac{1}{m_2}\right)
	\right]}{(p_1\cdot p_2)^2-m_1^2m_2^2}
\,,
\eeq
where $s=-(p_1+p_2)^2$.
Adding this expression, corresponding to the $s$-channel, to the one obtained from the $u$-channel yields in particular
\beq
\label{Box+CrossedBox-sub}
I_{\Box,s}^{(2s)}+I_{{\Box},u}^{(2s)}=
\frac{i\sqrt{\pi}\,\Gamma\left(\frac{5-D}{2}\right)\Gamma\left(\frac{D-3}{2}\right)^2}{2(4\pi)^{\frac{D}{2}}\Gamma(D-4)}
\left(\frac{q^2}{\hbar^2}\right)^{\frac{D-5}{2}}
\frac{m_1+m_2}{(p_1\cdot p_2)^2-m_1^2m_2^2}\,.
\eeq

As mentioned for the case of triangle integrals, we have focused on the soft-region expansion of box diagrams because it is the one containing terms with a non-analytic dependence on $q^2$ for generic $D$. The hard region, obtained expanding the original integral \eqref{boxq4regions} for $k\sim\mathcal O(1)$, namely $k\sim q_\perp,\,Q\gg q$, gives rise instead to terms with positive integer powers of $q^2$. For instance, the leading hard term for the box integral is given by
\beq
I^{(1h)}_\Box = \int \frac{d^Dk}{(2\pi\hbar)^{D}}\,\frac{\hbar^5}{(k^2)^2(k^2+(2Q-q_\perp)\cdot k)(k^2-q_\perp \cdot k)}
\eeq
so that, employing again Feynman parameters to rewrite the linear factors in the denominator in terms of a single one and using \eqref{formula12},
\beq\label{leadinghardh}
I^{(1h)}_\Box = \frac{i\Gamma\left(\tfrac{8-D}{2}\right)\Gamma\left(D-6\right)}{(4\pi)^{\tfrac{D}{2}}\Gamma(D-4)}\int_0^1 \frac{\hbar^{5-D}dx}{\left[-\left(xQ- \frac{q_\perp}{2}\right)^2-i\epsilon\right]^{\frac{8-D}{2}}}\,.
\eeq
This contribution is thus analytic in $q^2$ and finite in four dimensions. However, it is infrared divergent in, say, $D=5$. The box integral \eqref{boxq4regions} is however finite in five dimensions and this means that such a divergence must cancel out when adding the soft and the hard contributions: indeed, comparing \eqref{leadinghardh} with the subleading soft term \eqref{x+x-subbox} we see that the two divergent contributions cancel as $D\to5$ leading to a finite limit for $I^{(1h)}_\Box+I^{(2s)}_\Box$. 

\subsection{The potential region}
\label{PotReg}
Another region which can be useful for the expansion of Feynman integrals in the classical limit is the so-called potential region, as also argued in \cite{Cheung,Berndue}. To describe it, let us again consider the scalar triangle \eqref{triangle4regions}, which we write in the center-of-mass frame as
\beq
\label{triangle4potexp}
\begin{split}
	I_{\triangleright} = \int \frac{d^Dk}{(2\pi\hbar)^D}\,
	&
	\frac{\hbar^5}{(-(k^0)^2+|\vec k\,|^2-i\epsilon)
		(-(k^0)^2+|\vec k+\vec  q\,|^2-i\epsilon)}\\
	&\frac{1}{
		(-(k^0)^2+ |\vec k|^2 - 2E_1(p) k^0 + 2\vec p \cdot \vec k -i\epsilon)}\,,
\end{split}\eeq
where we have sent $k\to-k$ and adopted the same notation as in Section \ref{sect1}. 

As before, we are interested in the limit in which the transferred momentum $\vec q$ is of order $\hbar$ and is hence small with respect to the mass. We also consider the nonrelativistic limit, \emph{i.e.} the regime $|\vec p\,|\ll m_1$ in which the relative velocity $v$ is much smaller than the speed of light.
The potential region is then defined by the following scaling relations
\beq
k^0\sim  qv\,,\qquad
\vec k  \sim q\,,
\eeq
which break Lorentz invariance as they prescribe the time-component $k^0$ of the loop momentum to be negligible with respect to its spatial components $\vec k$.
The leading potential term is then obtained by simply neglecting the $(k^0)^2$ terms in the propagators, 
\beq\begin{split}
	I^{(1p)}_{\triangleright} = \int \frac{d^{D-1}\vec  k}{(2\pi\hbar)^{D-1}} \frac{\hbar^4}{|\vec k |^2	|\vec k+\vec   q\,|^2}\int \frac{dk^0}{2\pi}\,
	\frac{1}{
		(-2 E_1(p) k^0 + |\vec k|^2 + 2\vec p \cdot \vec k-i\epsilon)}\,.
\end{split}\eeq
The resulting integral over $dk^0$ is in principle ill defined, but can be evaluated by prescribing the application of the standard formula for the passage near a simple pole 
$
\frac{1}{x-i\epsilon}=\mathrm{PV}\,\frac{1}{x}+i\pi\delta(x)
$.
We thus obtain
\beq
I^{(1p)}_{\triangleright} = \frac{i}{4 E_1(p)} \int \frac{d^{D-1}\vec k}{(2\pi\hbar)^{D-1}}\,
\frac{\hbar^4}{|\vec k|^2	|\vec k+\vec  q\,|^2}\,.
\eeq
The remaining integral is elementary and can be evaluated by means of Feynman parameters, yielding
\beq
I_\triangleright^{(1p)}=\frac{i\sqrt\pi}{E_1(p)(4\pi)^{\frac{D}{2}}}\,\frac{\Gamma\left(\tfrac{D-3}{2}\right)^2\Gamma\left(\tfrac{5-D}{2}\right)}{2\Gamma(D-3)}\left(
\frac{q^2}{\hbar^2}
\right)^{\frac{D-5}{2}}\,.
\eeq
Taking into account the fact that $E_1(p)\approx m_1$ up to terms of order $v^2$ in the nonrelativistic limit,
this is the same as the leading soft result \eqref{scalarsoft}.

It would be interesting to reproduce the subleading soft term \eqref{scalarsoftsoft} from the subleading potential expansion, which is obtained from the higher-order terms in Taylor series of the integrand in \eqref{triangle4potexp} for small $(k^0)^2$. However, the resulting integral in $dk^0$ presents further difficulties, in particular due to appearance of a double pole.

Let us now turn to the potential-region expansion of the massive box \eqref{box4regions}. 
We go to the center-of-mass frame, adopting the same conventions as in Section~\ref{sect1},
so that
\begin{equation}\begin{split}
I_\Box &= \int \frac{d^Dk}{(2\pi\hbar)^{D}}\,
\frac{\hbar^5}{(-(k^0)^2+\vec k\,^2-i\epsilon)(-(k^0)^2+|\vec k-\vec  q\,|^2-i\epsilon)}\\
&\frac{1}{(-(k^0)^2+\vec k^2+2E_1k^0-2\vec  p  \cdot \vec  k-i\epsilon)(-(k^0)^2+\vec k^2-2E_2 k^0-2\vec  p  \cdot \vec k-i\epsilon)}\,.
\end{split}\end{equation}
In addition to the classical limit, which consists here in sending $\hbar\to0$ in such a way that
\beq
\vec q \sim \mathcal O(\hbar)\,,\qquad
\vec q_\perp \sim \mathcal O(1)\,,
\eeq 
where $\vec q_{\perp}=\vec p + \vec p\,'$, we also consider the nonrelativistic limit of small $v$, as we did for the triangle. We then adopt the scaling relations
\beq
k^0\sim qv \,\qquad 
\vec  k \sim q\,, 
\eeq 
which characterize the potential region for the loop momentum. We are thus justified in neglecting the $(k^0)^2$ appearing in the denominator, to leading order,
\beq
I_\Box^{(1p)}= \int \frac{d^Dk}{(2\pi\hbar)^{D}}\,
\frac{\hbar^5}{\vec k\,^2|\vec k-\vec  q\,|^2(2E_1k^0+\vec k^2-2\vec  p  \cdot \vec k-i\epsilon)(-2E_2 k^0+\vec k^2-2\vec  p  \cdot \vec k-i\epsilon)}\,.
\eeq
The integral in $dk^0$ can be performed with the help of the residue theorem, leading to
\beq
I_\Box^{(1p)}= \frac{i}{2E_p} \int \frac{d^{D-1}\vec k}{(2\pi\hbar)^{D-1}}\,
\frac{\hbar^4}{\vec k^2|\vec k-\vec  q\,|^2(\vec k^2-2\vec  p  \cdot \vec k-i\epsilon)}\,.
\eeq
Letting $\vec k \to \vec  p -\vec k$, we have 
\beq\label{boxpottobeexp}
I_\Box^{(1p)}= \frac{i}{2E_p} \int \frac{d^{D-1}\vec k}{(2\pi\hbar)^{D-1}}\,
\frac{\hbar^4}{|\vec k-\vec  p\, |^2|\vec k-\vec  p\, '|^2(\vec k^2-|\vec  p\,|^2-i\epsilon)}\,,
\eeq
so that we have reduced the problem to the evaluation of a Euclidan version of the triangle integral with an effective ``squared mass'' $m^2=-|\vec  p\,|^2-i\epsilon$. Indeed, with an appropriate choice of routing for the loop momentum, the triangle integral \eqref{triangle4regions} can be written as follows
\beq
I_{\triangleright} = i\int \frac{d^Dk_E}{(2\pi\hbar)^D}\,\frac{\hbar^5}{(k_E-p_{1E})^2(k_E-p _{3E})^2\left(k_E^2+\tfrac{m^2}{\hbar^2}\right)}\,,
\eeq
after Wick rotation, and therefore the above integral can be obtained from this one by the identifications 
\beq
D\to D-1\,,\qquad m\to -i\hbar|\vec  p\,|\,.
\eeq
Consequently, thanks to \eqref{scalarsoft} and \eqref{scalarsoftsoft}, we find
\begin{equation}
\begin{split}
\label{integrale}
&\int \frac{d^{D-1} \vec k}{(2\pi\hbar)^d}\frac{\hbar^4}{|\vec{k}-\vec{p}\,|^2|\vec{k}-\vec{p}\,'|^2(\vec{k}^2-{\vec{p}\,}^2-i \epsilon)}\\
= \, &
\frac{i\pi}{\hbar\, (4\pi)^{\frac{D}{2}}|\vec p |}\frac{\Gamma\left(\tfrac{6-D}{2}\right)\Gamma^2(\frac{D-4}{2})}{\Gamma(D-4)} \left(\frac{q^2}{\hbar^2}\right)^{\frac{D-6}{2}}
\!\!\!\!\!\!\!\!
+\frac{1}{2|\vec p\,|^2(4 \pi)^{\frac{D-1}{2}}}\frac{\Gamma\left(\tfrac{5-D}{2}\right)\Gamma^2(\frac{D-3}{2})}{\Gamma(D-4)} \left(\frac{q^2}{\hbar^2}\right)^{\frac{D-5}{2}}
\!\!\!\!\!\!\!\!+\cdots\ .
\end{split}
\end{equation}

We thus have, retaining the first two nontrivial orders for the soft-region expansion of \eqref{boxpottobeexp},
\beq\begin{split}
	I_\Box^{(1p)} &= -\frac{\pi}{\hbar|\vec p\,|E_p}
	\frac{\Gamma\left(\tfrac{D-4}{2}\right)^2\Gamma\left(\tfrac{6-D}{2}\right)}{2(4\pi)^{\tfrac{D}{2}}\Gamma(D-4)}
	\left(\frac{q^2}{\hbar^2}\right)^{\frac{D-6}{2}}
	\\
	&+ \frac{i\sqrt\pi}{|\vec  p\,|^2E_p}
	\frac{\Gamma\left(\tfrac{D-3}{2}\right)^2\Gamma\left(\tfrac{5-D}{2}\right)}{2(4\pi)^{\tfrac{D}{2}}\Gamma(D-4)}
	\left(\frac{q^2}{\hbar^2}\right)^{\frac{D-5}{2}}
	+\cdots
	\,.
\end{split}
\label{PotRegSoft}
\eeq

Note that the first line coincides with the leading order \eqref{Box+CrossedBox} for the  soft expansion of the sum of box and crossed box diagrams written in the center-of-mass frame, where $|\vec p\,|E_p=\sqrt{(p_1\cdot p_2)^2-m_1^2m_2^2}$.
Indeed, in the potential region, the crossed box diagram gives zero to leading order since the poles in $k^0$ both lie in the upper half plane.

However, the subleading order does not coincide with \eqref{Box+CrossedBox-sub}. It is in fact proportional to it, but instead of the total mass $m_1+m_2$ it displays a factor $E_p$, the center-of-mass energy, so that the two results do agree in the nonrelativistic limit $v\ll1$. This is in general to be expected, since the leading potential contribution $I_{\Box}^{(1p)}$ is only reliable to first order in the nonrelativistic limit. 
		
A more complete comparison between the results coming from the potential region and the ones obtained from the soft region for \emph{generic} velocities, \emph{i.e.} beyond the nonrelativistic regime, should be performed after resumming the potential series to all orders in $v$. However, the evaluation of subleading potential integrals is quite complicated due to the fact that they are in principle ill defined, as we have already seen for the triangle integral. A viable alternative to the evaluation of such integrals could be provided by an extension of the nonrelativistic integration techniques discussed in \cite{Berndue} to the case of generic dimensions.

In conclusion the potential region provides an expression for the non-analytic  terms terms in the small-$q$ expansion of the relevant Feynman integrals that agrees with the one furnished by the soft region at least to leading order in the nonrelativistic limit. In contrast, the soft region directly provides the non-analytic terms in the small-$q$ expansion in a fully relativistic manner. Let us also mention once more that the soft region gives rise to the needed cancellation of the spurious divergences appearing in the hard region, again without involving the nonrelativistic limit, as for instance between \eqref{x+x-subbox} and \eqref{leadinghardh} as $D\to5$.

\subsection{Auxiliary integrals}
\label{detailsss}
In this subsection we collect a number of useful standard techniques and results that allow one to explicitly evaluate the Feynman integrals presented above. To simplify the presentation, all quantities appearing in this section are understood to be dimensionless.
We first recall that, in $D$-dimensional Euclidean space, we have the general formula
\beq\label{Dimregintegral}
\int \frac{d^D\ell_E}{(2\pi)^D}\,\frac{(\ell_E^2)^\beta}{(\ell_E^2+\Delta_E^2)^{\alpha}}
=
\frac{\Gamma\left(\beta+\frac{D}{2}\right)\Gamma\left(\alpha-\beta-\frac{D}{2}\right)}{(4\pi)^\frac{D}{2}\Gamma\left(\alpha\right)\Gamma\left(\frac{D}{2}\right)}\,(\Delta_E^2)^{\frac{D}{2}-\alpha+\beta}\,.
\eeq

Let us consider 
\beq
I(p^2)=\int \,\frac{d^D\ell}{(\ell^2-i\epsilon)^{\lambda_1}(\ell^2-2p\cdot \ell-i\epsilon)^{\lambda_2}}\,,
\eeq
where $p^\mu$ is a time--like vector, $(-p^2)>0$. Introducing Feynman parameters we have
\beq
I(p^2)=\frac{\Gamma(\lambda_1+\lambda_2)}{\Gamma(\lambda_1)\Gamma(\lambda_2)}\int_0^1dx\,(1-x)^{\lambda_1-1}x^{\lambda_2-1} \int \frac{d^D\ell}{(\ell^2-2xp\cdot \ell-i\epsilon)^{\lambda_1+\lambda_2}}\,.
\eeq
Shifting $\ell$ by $xp$ so as to complete the square in the denominator, performing the Wick rotation $(\ell^0,\vec  \ell\ ) = (i\ell^0_E, \vec  \ell_E)$ and employing equation \eqref{Dimregintegral}, one then obtains
\beq
I(p^2) =
i\pi^{\frac{D}{2}} \frac{\Gamma(\lambda_1+\lambda_2-\frac{D}{2})}{\Gamma(\lambda_1)\Gamma(\lambda_2)} \int_0^1 (1-x)^{\lambda_1-1}x^{D-2\lambda_1-\lambda_2-1}dx\,(-p^2)^{\frac{D}{2}-\lambda_1-\lambda_2}\,.
\eeq
Finally, recognizing the Beta function appearing in the last equation, we get the formula (cf. \cite[eq. (A.13)]{Smirnov})
\beq\label{formula12}
\int \,\frac{d^D\ell}{(\ell^2-i\epsilon)^{\lambda_1}(\ell^2-2p\cdot \ell-i\epsilon)^{\lambda_2}} = i\pi^{\frac{D}{2}}\ \frac{\Gamma(\lambda_1+\lambda_2-\frac{D}{2})\Gamma(D-2\lambda_1-\lambda_2)}{\Gamma(\lambda_2)\Gamma(D-\lambda_1-\lambda_2)(-p^2)^{\lambda_1+\lambda_2-\frac{D}{2}}}\,.
\eeq
In a very similar way, one can also derive (cf. \cite[eq. (A.7)]{Smirnov})
\beq\label{ellell-q}
\int \frac{d^D\ell}{(\ell^2-i\epsilon)^{\lambda_1}\left((\ell-q)^2-i\epsilon\right)^{\lambda_2}} = i\pi^{\frac{D}{2}}\, \frac{\Gamma\left(\lambda_1+\lambda_2-\tfrac{D}{2}\right)\Gamma\left(\tfrac{D}{2}-\lambda_1\right)\Gamma\left(\tfrac{D}{2}-\lambda_2\right)}{\Gamma(\lambda_1)\Gamma(\lambda_2)\Gamma(D-\lambda_1-\lambda_2)(q^2)^{\lambda_1+\lambda_2-\frac{D}{2}}}\,.
\eeq
Let us now consider the following integral
\beq
I_\perp(q^2,r^2)=\int\frac{d^D\ell}{(\ell^2-i\epsilon)^{\lambda_1}((q-\ell)^2-i\epsilon)^{\lambda_2}(2r\cdot \ell-i\epsilon)^{\lambda_3}}\,,
\eeq
where $r^\mu$ is time--like, $(-r^2)>0$, and $q\cdot r=0$, so that $q^\mu$ is space--like, $q^2>0$. Proceeding as in the previous case, we obtain
\beq\begin{aligned}
I_\perp(q^2,r^2)&=
i\pi^{\frac{D}{2}}\,
\frac{\Gamma(\lambda_1+\lambda_2+\lambda_3-\frac{D}{2})}{\Gamma(\lambda_1)\Gamma(\lambda_2)\Gamma(\lambda_3)}
\int_0^\infty dx\,x^{\lambda_1-1}\int_0^\infty dy\, y^{\lambda_2-1}\int_0^\infty dz\,z^{\lambda_3-1}\,\\
&\times\delta(1-x-y-z)\frac{(z^2(-r^2)+xy\,q^2)^{\frac{D}{2}-\lambda_1-\lambda_2-\lambda_3}}{(x+y)^{D-\lambda_1-\lambda_2-\lambda_3}}
\,,
\end{aligned}\eeq
where $x$, $y$ and $z$ are Feynman parameters.
We change variables according to 
\beq
x=\lambda x_1 \sqrt{\frac{(-r^2)}{q^2}}\,,\qquad y=\lambda x_2 \sqrt{\frac{(-r^2)}{q^2}}\,,\qquad z=\lambda\,,
\eeq
which simplifies the integral to
\beq
I_\perp(r^2,q^2)=i\pi^{\frac{D}{2}}\,
\frac{\Gamma(\lambda_1+\lambda_2+\lambda_3-\frac{D}{2})}{\Gamma(\lambda_1)\Gamma(\lambda_2)\Gamma(\lambda_3)}\frac{I'}{(q^2)^{\lambda_1+\lambda_2+\frac{\lambda_3-D}{2}}(-r^2)^{\frac{\lambda_3}{2}}}\,,
\eeq
where $I'$ is an integral which does not depend on $q^2$ nor on $r^2$,
\beq
I'=
\int_0^\infty dx_1\,x_1^{\lambda_1-1} \int_0^\infty dx_2\,x_2^{\lambda_2-1}\, \frac{(1+x_1x_2)^{\frac{D}{2}-\lambda_1-\lambda_2-\lambda_3}}{(x_1+x_2)^{D-\lambda_1-\lambda_2-\lambda_3}}\,.
\eeq
This can be evaluated performing the substitution $x_1 = uv$ and $x_2 = \frac{u}{v}$, which factorizes it into two integrals of the type
\beq
\int_0^\infty u^\alpha (1+u^2)^{\beta} du = \frac{\Gamma(-\frac{\alpha+2\beta+1}{2})\Gamma(\frac{\alpha+1}{2})}{2\Gamma(-\beta)}\,,
\eeq
conveniently evaluated letting $x=\frac{1}{1+u^2}$. 

In conclusion, for the two orthogonal vectors $q\cdot r=0$, we obtain (cf. \cite[eq. (A.27)]{Smirnov}) 
\beq\begin{aligned}\label{formulaperp}
I_\perp(q^2,r^2)&=\int \frac{d^D\ell}{(\ell^2-i\epsilon)^{\lambda_1}((q-\ell)^2-i\epsilon)^{\lambda_2}(2r\cdot \ell-i\epsilon)^{\lambda_3}}\\
&=i\pi^{\frac{D}{2}}\ \frac{\Gamma(\lambda_1+\lambda_2+\frac{\lambda_3-D}{2})\Gamma(\frac{\lambda_3}{2})}{2\Gamma(\lambda_1)\Gamma(\lambda_2)\Gamma(\lambda_3)\Gamma(D-\lambda_1-\lambda_2-\lambda_3)}
\ \frac{\Gamma(\frac{D-\lambda_3}{2}-\lambda_1)\Gamma(\frac{D-\lambda_3}{2}-\lambda_2)}{(q^2)^{\lambda_1+\lambda_2+\frac{\lambda_3-D}{2}}(-r^2)^{\frac{\lambda_3}{2}}}\,.
\end{aligned}\eeq
Variants of the above integral that can be evaluated in a similar fashion, still under the assumption $q\cdot r=0$, are
\begin{equation}
\begin{aligned}\label{formulaperp2}
&I_\perp^{(1)}(q^2,r^2)=\int \frac{(q\cdot \ell)\,d^D\ell}{(\ell^2-i\epsilon)^{\lambda_1}((q-\ell)^2-i\epsilon)^{\lambda_2}(2r\cdot \ell-i\epsilon)^{\lambda_3}}\\
&=i\pi^{\frac{D}{2}}\ \frac{\Gamma(\lambda_1+\lambda_2+\frac{\lambda_3-D}{2})\Gamma(\frac{\lambda_3}{2})}{2\Gamma(\lambda_1)\Gamma(\lambda_2)\Gamma(\lambda_3)\Gamma(D-\lambda_1-\lambda_2-\lambda_3+1)}
\ \frac{\Gamma(\frac{D-\lambda_3}{2}-\lambda_2)\Gamma(\frac{D-\lambda_3}{2}-\lambda_1+1)}{(q^2)^{\lambda_1+\lambda_2+\frac{\lambda_3-D}{2}-1}(-r^2)^{\frac{\lambda_3}{2}}}
\end{aligned}
\end{equation}
and
\begin{equation}
\begin{aligned}\label{formulaperp3}
&I_\perp^{(2)}(q^2,r^2)=\int \frac{(q\cdot \ell)^2\,d^D\ell}{(\ell^2-i\epsilon)^{\lambda_1}((q-\ell)^2-i\epsilon)^{\lambda_2}(2r\cdot \ell-i\epsilon)^{\lambda_3}}\\
&=i\pi^{\frac{D}{2}}\ \frac{\Gamma(\lambda_1+\lambda_2+\frac{\lambda_3-D}{2})\Gamma(\frac{\lambda_3}{2})}{2\Gamma(\lambda_1)\Gamma(\lambda_2)\Gamma(\lambda_3)\Gamma(D-\lambda_1-\lambda_2-\lambda_3+2)}
\ \frac{\Gamma(\frac{D-\lambda_3}{2}-\lambda_1+1)\Gamma(\frac{D-\lambda_3}{2}-\lambda_2+1)}{(q^2)^{\lambda_1+\lambda_2+\frac{\lambda_3-D}{2}-2}(-r^2)^{\frac{\lambda_3}{2}}}\\
&\times \left(
\frac{D-2\lambda_1-\lambda_3+2}{D-2\lambda_2-\lambda_3}-
\frac{1}{D+2-2\lambda_1-2\lambda_2-\lambda_3}
\right)\,.
\end{aligned}
\end{equation}


%

\end{document}